\documentclass[12pt]{iopart}

\usepackage{graphicx} 

\expandafter\let\csname equation*\endcsname\relax
\expandafter\let\csname endequation*\endcsname\relax
\usepackage{amsmath}

\begin{document}

\title[Time-dependent Correlation Functions in Open Quadratic Fermionic Systems]{Time-dependent Correlation Functions in Open Quadratic Fermionic Systems}

\author{Pavel Kos and Toma\v{z} Prosen}

\address{Department of physics, FMF, University of Ljubljana, Jadranska 19, SI-1000
Ljubljana, Slovenia}
\vspace{10pt}
\begin{indented}
\item[]June 2017
\end{indented}

\begin{abstract}
We formulate and discuss explicit computation of dynamic correlation functions in open quadradic fermionic systems which are driven and dissipated by the Lindblad jump processes that are linear in canonical fermionic operators.
Dynamic correlators are interpreted in terms of local quantum quench where the pre-quench state is the non-equilibrium steady state, i.e. a fixed point of the Liouvillian.
As an example we study the XY spin 1/2 chain and the Kitaev Majorana chains with boundary Lindblad driving, whose dynamics exhibits asymmetric (skewed) light cone behaviour. We also numerically treat the two dimensional XY model and the XY spin chain with additional Dzyaloshinskii-Moriya interactions. The latter exhibits a new non-equilibrium phase transition which can be understood in terms of bifurcations of the quasi-particle dispersion relation.

Finally, considering in some detail the periodic Kitaev chain (fermionic ring) with dissipation at a single (arbitrary) site, we present analytical expressions for the first order corrections (in the strength of dissipation) to the spectrum and the non-equilibrium steady state (NESS) correlation functions.
\end{abstract}

%
%
%
%
%
\maketitle

{
}

\section{Introduction}

The motivation for studying open quantum systems, where a finite but possibly large quantum system interacts with an infinite environment, is twofold: on the one hand it is a way of attacking the fundamental questions about the quantum mechanics such as the quantum measurement problem, the decoherence and the dynamics far from equilibrium.
On the other hand, it is very relevant for  experimental quantum physics, which has advanced a lot over the last few years and is capable of having a good control over the many body states and their dynamics (e.g. in ultracold gases, trapped ions, etc).

The decoherence is one of the main players in the second quantum revolution \cite{Dowling1655}, which will develop new quantum technologies. It plays an important role in quantum computing, state preparation and quantum memories.
Furthermore, studying non-equilibrium steady states (NESS) of open systems give us new interesting phases of matter separated by non-equilibrium phase transitions \cite{Prosen2010a, PhysRevLett.107.060403,PhysRevA.87.012108,mitra}.

The dynamics far from equilibrium is still poorly understood even in the context of integrtable systems. In recent years there has been much progress on understanding dynamics following an instantaneous local or global quantum quench \cite{PhysRevLett.113.187203,PhysRevLett.108.077206,Caux,CalabreseCardy,EsslerFagotti} in closed systems. In the context of open systems, even though some progress in understanding the structure of non-equilibrium steady states of boundary driven integrable systems has been achieved \cite{TP15}, the relaxation (Liouvillian) dynamics is still essentially unexplored. 

In this paper we propose to study the simplest of such problems. Specifically, we consider a protocol in which we start in a non-equilibrium steady state, perform a local quench and then study the relaxation dynamics in terms of local observables. In this way we define Liouvillian dynamic correlation functions and set up a procedure of calculating them for quasi-free fermionic many-body systems.

First we discuss a general quadratic one dimensional fermionic model that is governed by the Lindblad master equation with general linear Lindblad operators. 
We investigate its spectrum and the static and the \textit{dynamic} correlation functions (section 2). In the second part, we focus on a specific problem given by a weakly coupled fermionic ring with superconducting pairing terms (sections 3), which can be as well applied to study boundary driven $XY$ spin chain model. 
In section 4 we provide some interesting numerical results for static and dynamic NESS correlation functions of the one-dimensional and two-dimensional quadratic fermionic models. In particular, in the boundary driven XY chain we observe skewed (asymmetric) light-cone behaviour, with the real part of two-point correlations with the points at late times to the left (right) well approximated by the thermal correlators of the right (left) bath. Studying  boundary driven $XY$ spin chain with the presence of Dzaloshinsky-Moriya interaction terms, we observe a rich non-equilibrium phase diagram with additional critical lines which correspond to bifurcations of quasi-particle dispersion relation.
In section 5, we then focus on the perturbative/asymptotic analysis of the Liouvillian spectrum and static NESS correlator of the Kitaev ring in the presence of weak dissipation/driving. 
Most interestingly, with the dissipative driving at just one site we encounter a unique steady state, where dissipation creates entanglement/quantum correlations between opposite pairs of fermions/spins over large distances.

\section{Statement of the Fermionic Problem}
%
The evolution of an open system's density matrix $\rho(t)$ is given by the Lindblad master equation \cite{lindblad1976}, which is the most general Markovian, trace preserving and completely positive non-unitary map:
\begin{align}
    \frac{\text{d}\rho}{\text{d}t}= \hat{\mathcal{L}} \rho:=-{\rm i}[H,\rho]+\sum_{\mu=1}^M \big( 2 L_\mu \rho L_\mu^\dagger- \{L_\mu^\dagger L_\mu,\rho \} \big).
\end{align}
We focus on quadratic fermionic Hamiltonians $H$ for systems of length $n$, written in terms of $2n$ Majorana fermions $w_j$ (we use a convention $w^2_j=1$) combined together with a $2n \times 2n$ anti-symmetric matrix $\textbf{H}$  and write the Hamiltonian as $H=\sum_{j,k=1}^{2n} w_j \textbf{H}_{jk} w_k$. The linear Lindblad operators are written with the a help of $2n$-component vectors $\underline{l_{\mu}}$ as $L_\mu= \sum_j l_{\mu,j} w_j$.

\subsection{Third Quantization}
A convenient general procedure for treating such problems is the so-called third quantization \cite{Prosen2008,Prosen2010,Prosen_3_bosons,Guo2017}, that is based on the quantization in the Fock space of operators. Here we quickly review it and fix the notation. See also Ref.~\cite{kosov} for an alternative but equivalent {\em super-fermion} formulation.

We are interested in the dynamics of a density matrix $\rho(t)$ which is a non-negative unit trace operator. Vectorizing the space of operators, the latter are now expressed as vectors in a $4^n$ dimensional Fock space $\mathcal{K}$ of operators (Liouville-Fock space) and the superoperators become operators over $\mathcal{K}$.  We select an orthonormal canonical basis for $\mathcal{K}$ as:
\begin{align}
    P_{\alpha_1,\dots,\alpha_{2n}} :=2^{-n/2} w_1^{\alpha_1} w_2^{\alpha_2} \dots w_{2n}^{\alpha_{2n}} , && \alpha_j \in \{ 0,1\},
\end{align}
and introduce the creation and the annihilation linear operators $\hat{c}_j$, $\hat{c}_j^\dagger$ over $\mathcal{K}$: 
$\hat{c}_j^\dagger | P_{\underline{\alpha}} \rangle =\delta_{\alpha_j,0} |w_j P_{\underline{\alpha}} \rangle $ and
$\hat{c}_j | P_{\underline{\alpha}} \rangle =\delta_{\alpha_j,1} |w_j P_{\underline{\alpha}} \rangle $, which obey the canonical anticommutation relations (CAR). 
They are combined to the $4 n$ Majorana fermionic super-operators over $\mathcal{K}$:
\begin{align}
    \hat{a}_{1,j}:=\frac{1}{\sqrt{2}}(\hat{c}_j+\hat{c}_j^\dagger), && \hat{a}_{2,j}:=\frac{{\rm i}}{\sqrt{2}}(\hat{c}_j-\hat{c}_j^\dagger),
\end{align}
obeying CAR $\{\hat{a}_{\nu,j},\hat{a}_{\nu',j'}\} =  \delta_{\nu,\nu'}\delta_{j,j'}$, with which we can express the Liouvillian map $\hat{\mathcal{L}}$ simply as:
\begin{align}
\frac{d}{dt}|\rho(t) \rangle& =\hat{\mathcal{L}}|\rho(t) \rangle, &&
    \hat{\mathcal{L}}= \underline{\hat{a}} \textbf{A} \underline{\hat{a}} -A_0 \textbf{1}_{4n}.
\end{align}

The constant $A_0$ is unimportant \cite{Prosen2008}, and the $4n \times 4n$ matrix $\textbf{A}$ can be expressed in terms of the Hamiltonian matrix $\textbf{H}$ and the positive semidefinite Hermitian Lindblad bath matrix $\textbf{M}= \sum_\mu \underline{l_\mu} \otimes \underline{l_\mu}^*$. The matrices $\textbf{M}_r$ and $\textbf{M}_i$ are its real and imaginary parts.
\begin{align}
    \textbf{A}&=\begin{pmatrix}
    \textbf{X} & 2 {\rm i} \textbf{M} \\
    -2 {\rm i} \textbf{M}^T & \textbf{X}-4 {\rm i} \textbf{M}_i
    \end{pmatrix},&&
    \textbf{X}=-2 {\rm i} \textbf{H} + 2 \textbf{M}_r.
    \label{Xmatrix}
\end{align}
The matrix $\textbf{A}$ is unitarily equivalent to block triangular matrix $\tilde{\textbf{A}}=\textbf{U}\textbf{A}\textbf{U}^\dagger$,
\begin{align}
    \tilde{\textbf{A}}=\begin{pmatrix} -\textbf{X}^T & 4 {\rm i} \textbf{M}_i \\ 0& \textbf{X}  \end{pmatrix}, && 
    \textbf{U}=\frac{1}{\sqrt{2}}\begin{pmatrix} 1&-{\rm i}\\1& {\rm i} \end{pmatrix} \otimes \textbf{1}_{2n},
\end{align}
therefore its spectrum is determined by the eigenvalues $\{ \Lambda_j \}_{j=1}^{2n}$ of the $2n \times 2n$ non-Hermitian matrix $ \textbf{X}$.
The imaginary part of these eigenvalues is mostly determined by the Hamiltonian, whereas the real part is positive semi-definite and is a consequence of the dissipation encoded in $\textbf{M}_r$.
The spectrum of the Liouvillian operator $\hat{\mathcal{L}}$ consists of $4^n$ elements $\{ \sum_{\underline{\nu} \in \{0,1\}^{\otimes 2n}} \nu_k \Lambda_k \}$ and does not depend on $\textbf{M}_i$.

We could proceed by constructing ``non-unitary Bogoliubov transformation" diagonalising $\textbf{A}$ and express the Liouvillian operator in terms of the normal master modes
$\hat{\mathcal{L}}=-2 \sum_{j=1}^{2n}\Lambda_j \hat{b}_j' \hat{b}_j$,
where $\Lambda_j$ are the eigenvalues of $\textbf{X}$ and CAR obeying operators $\hat{b}_j'$ and  $\hat{b}_j$ are the linear combinations of the Majorana operators $\hat{a}$ \cite{Prosen2008, Prosen2010,Prosen2008a}. 
But this is not necessary, as will become clear in the next subsection.

\subsection{Correlation Functions}
\label{sec:Correlation functions}

The equal time two point correlation functions in a non-equilibrium steady state (NESS) can be calculated using the normal master modes, or directly by writing the dynamics of the two point correlations and calculating the fixed point \cite{Prosen2010,Prosen2010a,Zunkovic2010}. Since the Hamiltonian is quadratic and the coupling Lindblad operators $L_\mu$ are linear in fermionic operators, the NESS is a Gaussian state and the higher order correlation functions can be calculated from the two point correlation functions using the Wick contractions.


The two point correlation functions are given in terms of a covariance (or correlation) matrix $\textbf{C}$ \cite{Prosen2010,Prosen2010a,Zunkovic2010}:
 \begin{align}
      \textbf{C}_{jk}(t)= \Tr w_j w_k \rho(t)- \delta_{j,k}=2 \langle 1|  \hat{a}_{1,j} \hat{a}_{1,k}|\rho(t) \rangle - \delta_{j,k}. 
 \end{align}

To express the dynamics, we focus on the time derivative of the two point correlation functions. We adapt the super-Heisenberg picture and write the time derivative in terms of the time-dependent Majorana super-operators $\hat{a}_{\mu,z}(t)=e^{-\hat{\mathcal{L}}t} \hat{a}_{\mu,z} e^{\hat{\mathcal{L}}t}$
\begin{align}
 \frac{d \textbf{C}_{jk}(t)}{d t}&=\langle \frac{d w_j(t)}{d t} w_k(t) \rangle_{\rho(0)}
 + \langle w_j(t) \frac{d w_k(t)}{dt} \rangle_{\rho(0)}\\
 &=2 \langle 1| \frac{d \hat{a}_{1,j}(t)}{dt} \hat{a}_{1,k}(t)+\hat{a}_{1,j}(t) \frac{d \hat{a}_{1,k}(t)}{dt}|\rho(0) \rangle.
\end{align} 
The time derivative of the Majorana super-operators is:
\begin{align}
    \frac{d \hat{a}_{1,z}(t)}{dt}=[\hat{a}_{1,z}(t), \hat{\mathcal{L}}]= 2 \sum_s^{2n} \big( \textbf{A}_{(1,z),(1,s)} \hat{a}_{1,s} +\textbf{A}_{(1,z),(2,s)} \hat{a}_{2,s} \big).
    \label{eq:majorana der}
\end{align}
After simple manipulations we obtain the continuous Lyapunov equation that governs the time evolution of the equal-time correlation functions:
\begin{align}
-\frac{1}{2}\frac{d\textbf{C}(t)}{dt}=  \textbf{X}^T \textbf{C}(t)+ \textbf{C}(t) \textbf{X}- 4 {\rm i} \textbf{M}_i.
    \label{eq:Lyapunov}
\end{align} 
We have used $\Tr \rho(0)=1$, but not that $\rho(0)$ is self-adjoint. The time evolution of the correlation functions is closed and Markovian, i.e. it does no depend on the whole history of $\rho(t')$ but only on instantaneous covariance matrix $\textbf{C}(t)$.

\subsubsection{Correlations in a NESS.}
The two point correlations in a NESS $\textbf{C}(\infty)=\textbf{C}^\infty $ are determined by the condition $\frac{d\textbf{C}^\infty}{ dt}=0$. The NESS is unique if the eigenvalues of $\textbf{X}$ have strictly positive real part \cite{Prosen2008}. Moreover, there are more general theorems stating necessary conditions for uniqueness of the NESS \cite{Evans1977}, or non-uniqueness in the presence of strong symmetry \cite{Berislav}.
We can calculate $\textbf{C}^\infty$ using the equation \eqref{eq:Lyapunov}, $\frac{d\textbf{C}^\infty}{ dt}=0$ and numerical recipes for solving the continuous Lyapunov equation in $\mathcal{O}(n^3)$, for instance using a Bartels Stewart Algorithm \cite{NumLyap}.

For the analytical solution it is useful to use the spectral decomposition of the matrix $\textbf{X}$: $\textbf{X} \tilde{\textbf{P}}= \tilde{\textbf{P}} \Lambda$, $\Lambda=\text{diag}(\dots,\Lambda_j,\dots)$. The correlations in the NESS can be expressed as:
 \begin{align}
     \{ \Lambda, \tilde{\textbf{P}}^T \textbf{C}^\infty  \tilde{\textbf{P}} \}&=  4 {\rm i} \tilde{\textbf{P}}^T \textbf{M}_i  \tilde{\textbf{P}}, &
    ( \tilde{\textbf{P}}^T \textbf{C}^\infty  \tilde{\textbf{P}})_{kl}&=4 {\rm i}\frac{(\tilde{\textbf{P}}^T \textbf{M}_i  \tilde{\textbf{P}})_{kl}}{\Lambda_k+\Lambda_l}.
    \label{eq:corrNESS}
 \end{align}
Having computed $\tilde{\textbf{P}}^T \textbf{C}^\infty \tilde{\textbf{P}}$ from the equation \eqref{eq:corrNESS}, we can express the correlations in the NESS by $\textbf{C}^\infty=\tilde{\textbf{P}}^{-T}(\tilde{\textbf{P}}^T \textbf{C}^\infty \tilde{\textbf{P}}) \tilde{\textbf{P}}^{-1}$. In the right part of the equation \eqref{eq:corrNESS}, we see that the driving encoded in $\textbf{M}_i$ sets the correlations in the NESS. 
Note that it does not appear anywhere else in our derivation.
 
\subsubsection{Dynamics of the Correlations.}
The main goal of this paper is to study the dynamic response of non-equilibrium open systems. 
Writing the correlation matrix as $\textbf{C}(t)=\textbf{C}^\infty + (\textbf{C}(t)-\textbf{C}^\infty)$ and inserting it in the equation \eqref{eq:Lyapunov}, we see that the general solution for the correlation matrix is:
 \begin{align}
     \textbf{C}(t)= e^{-2t \textbf{X}^T} (\textbf{C}(0)-\textbf{C}^\infty) e^{-2t \textbf{X}}+\textbf{C}^\infty, 
     && \textbf{C}(0)_{j,m}= \Tr w_j w_m \rho(0)-\delta_{j,m}.
     \label{eq:dynamics1}
 \end{align}
 
 Using the diagonalization $\textbf{X} \tilde{\textbf{P}}= \tilde{\textbf{P}} \Lambda$ 
  and definition $\textbf{G} (t)=\tilde{\textbf{P}}^T (\textbf{C}(t)-\textbf{C}^\infty) \tilde{\textbf{P}}$ the simple evolution of the correlations in the new basis has the form:
 \begin{align}
    \textbf{G}(t)_{j,m}=e^{-2t (\Lambda_j+\Lambda_m)} \textbf{G}(0)_{j,m}.
    \label{eq:dynamics2}
\end{align}

The equations \eqref{eq:dynamics1} and \eqref{eq:dynamics2} are very general and give us important information about the dynamics in quadratic open quantum systems. The time-dependent correlation functions are thus given in terms of simple propagation of the initial condition.

The positive real part of the spectrum of matrix $\textbf{X}$, which is mostly determined by dissipation encoded in $\textbf{M}_r$, determines the decay of the correlations in NESS. 
The smallest real part of $\Lambda_j$, i.e. the Liouvillian gap, determines the longest possible time scale of the decay.
The imaginary part of the spectrum is mostly determined by the Hamiltonian $\textbf{H}$ and produces oscillations and the light cone like effect which will be discussed later. 
The driving encoded in the $\textbf{M}_i$ influences the correlations in the NESS $\textbf{C}^\infty$.

In order to investigate the dynamics we would like to calculate the time-dependent correlation functions for specific simple non-stationary initial states. To avoid breaking the super-parity \cite{Prosen2010a} we are interested in the four-point time-dependent correlation functions of the form
\begin{align}
    \Tr \left( (w_j w_m)(t) (w_k w_l)(0) \rho \right),
    \label{eq:4pt Corr}
\end{align}
which captures for example the time-dependent density-density correlations  and  spin-spin correlator $ \langle \sigma^z_k(t) \sigma^z_l(0)\rangle$ for the spin systems after performing the Jordan-Wigner transformation.
%
%
It is especially interesting and simple to look at the case when $\rho=\rho_{\rm NESS}$ is the NESS density matrix. 
In the derivation of the map $\textbf{C}(0)\to \textbf{C}(t)$, Eqs. (\ref{eq:dynamics1},\ref{eq:dynamics2}), we have never required $\rho(0)$ to be a valid density matrix (i.e. non-negative Hermitian operator), in fact we can use any element of the Fock space ${\cal K}^+$ with positive super-parity (see Ref. \cite{Prosen2008} for definitions).
In order to calculate the four point correlation function \eqref{eq:4pt Corr}, we take the initial condition $\rho(0) =w_k w_l \rho_{\rm NESS}/ \Tr(w_k w_l \rho_{\rm NESS})$, for some fixed indices $k,l$,
%
so the initial value correlation matrix reads:
\begin{align}
    \textbf{C}(0)_{j,m}+\delta_{j,m}&=\Tr w_j w_m \rho(0) 
    =\langle w_j w_m w_k w_l  \rangle / \langle
    w_k w_l  \rangle 
    \nonumber\\
    &= \langle  w_j w_m \rangle +
    (\langle  w_j w_l  \rangle \langle  w_m w_k  \rangle -\langle  w_j w_k  \rangle \langle  w_m w_l  \rangle )/\langle  w_k w_l \rangle,
    \label{eq:initial}
\end{align}
where we have used a short hand notation for expectation values with respect to NESS, $\langle A\rangle\equiv \Tr (A \rho_{\rm NESS})$. The solution of this problem can be also seen as a Green's function of the time evolution, and can be used to study any initial condition. Another important initial condition is the case of a thermal initial state, where the initial condition is given by Eq. \eqref{eq:initial} where we take expectation values with respect to $\rho=Z^{-1}\exp(-\beta H)$.

\subsection{Local Open Quantum Quench}
There was a lot of interest recently in studying global quantum quenches. This naturally lead also to the interest in the local quantum quenches, where one starts with a ground state of a Hamiltonian that is only locally different from the post-quench Hamiltonian (see e.g. \cite{CalabreseCardy,EsslerFagotti}). 
Here we look at the local quench from a different, open system's perspective and define it via the following protocol:
\begin{itemize}
\item[1)] Wait for the open quantum system to relax into the steady state (NESS).
\item[2)] Perform a local quench by measuring a local observable (e.g., $n_k=c_k^\dagger c_k$).
\item[3)] Use the resulting density matrix as the initial condition.
\item[4)] Measure the correlations in the state at later times.
\end{itemize}

Let us look at an example where we measure a particle number $n_k$ at site $k$ at time $t=0$ in NESS, and continue with the experiment provided we measured $n_k=1$. The initial condition is then given by the density matrix
\begin{align}
  \rho(0) =&\frac{P \rho_{\rm NESS} P}{\Tr P \rho_{\rm NESS} P}, & P&=c^\dagger_k c_k= \frac{1-{\rm i} w_{2k-1} w_{2k}}{2},
\end{align}
  where $P$ is a projector on the subspace with a particle at site $k$. Explicitly, $\rho(0)$ and the initial condition $\textbf{C}(0)$ for the later time correlation functions are given by:
  \begin{align}
  \rho(0) =&\frac{1}{2}\frac{ \rho-  w_{2k-1} w_{2k}\rho w_{2k-1} w_{2k} -{\rm i}  w_{2k-1} w_{2k}\rho-{\rm i} \rho  w_{2k-1} w_{2k} }{1-{\rm i} \langle w_{2k-1} w_{2k} \rangle},\\
  \textbf{C}(0)_{j,m}=&\frac{ \langle w_j w_m w_{2k-1} w_{2k} \rangle}{{\rm i}+ \langle w_{2k-1} w_{2k} \rangle},\qquad\textbf{C}(0)_{j,j}=0.
\end{align}
The initial condition can be considered to be a special case of \eqref{eq:initial} up to the normalization, so we will focus later on calculating the time-dependent correlation functions resulting from local quantum quenches.

\section{Application to One Dimensional Quadratic Fermionic Chain}

In the second part of this paper, we will focus on the quadratic fermionic Hamiltonian with superconductive pairing
\begin{align}
\begin{split}
     H &= \sum_m (c_m^\dagger c_{m+1}+ \gamma c_m^\dagger c_{m+1}^\dagger + h. c. ) + h \ (2c_m^\dagger c_m-1) \\
     &=- {\rm i} \sum_m \left(\frac{1+\gamma}{2} w_{2m} w_{2m+1} -\frac{1-\gamma}{2} w_{2m-1} w_{2m+2} + h \ w_{2m-1} w_{2m} \right) ,
     \end{split}
     \label{eq:Hami}
\end{align}
which corresponds to the Heisenberg XY spin chain after the Jordan-Wigner transformation for open-boundary spin chain. This problem is both rich and treatable, which can be seen from the huge number of works treating it. It is also well known as the Kitaev Majorana chain \cite{1063-7869-44-10S-S29}. The main focus will be on the periodic boundary conditions, but we will also numerically treat the open (free) boundary conditions (relevant for the spin problem).

Coupling with the environment is parametrized by the following bath (Lindblad) operators $L_\mu$, which create/destroy fermioinic particle at site $j$
\footnote{The Jordan-Wigner transformation complicates the expressions for the spin systems, where we can couple to the bath only the first and the last spin site in order to keep the quadratic structure of the problem.
}:
\begin{align}
L_+^j&= \sqrt{\Gamma_{+}^{j}} c^\dagger_j,
&
L_-^j&= \sqrt{\Gamma_{-}^{j}} c_j,
&
         L_{\pm}^{j}&=\frac{1}{2} \sqrt{\Gamma_{\pm}^{j}} (w_{2j-1} \mp {\rm i} w_{2j}), 
         \label{eq:driving}
\end{align}
where $\Gamma^j_\pm$ are positive coupling constants. In the XY model, they are related to baths' temperatures/magnetizations of the non-interacting spins as $ \Gamma_-/\Gamma_+= e^{-2 h \beta}$ \cite{Prosen2008}. 
For easier comparison with the thermal closed system, we parametrise these constants in a slightly different way, in terms of new parameters $S_j,\beta_j$:
\begin{align}
    \Gamma_\pm^{j}= \frac{S_j e^{\pm h \beta_{j}}}{ 2 \cosh h \beta_{j}},
    && \Gamma_+^{j}+\Gamma_-^{j}=S_j,
    && \Gamma_+^{j} - \Gamma_-^{j}=S_j \tanh h \beta_j.
\end{align}

\subsection{Application to the XY Spin Chain}
The XY spin chain has attracted a lot of interest in the past (see e.g., Refs. \cite{Lieb1961,Mazur1973,Niemeijer1967}), since it is a prime example of a solvable system. 
Recently, it was reviewed in the quantum information community, where it is a popular example for understanding the entanglement generation  \cite{PhysRevE.89.022102,PhysRevA.78.010306,PhysRevLett.88.107901,PhysRevA.80.032304}, the quantum speed limit \cite{Hou2017,Wei2016} and other quantum information concepts.
It was also a prime example for the open systems treated with the method of third quantization (see e.g. \cite{Prosen2008, Prosen2008a, PhysRevE.89.022102}), the dynamic response was computed in \cite{PhysRevA.93.032101}.
The XX chain was solved directly \cite{1742-5468-2010-05-L05002,PhysRevE.83.011108} and its transport properties were studied in \cite{PhysRevE.84.051115}.
The Hamiltonian \eqref{eq:Hami} is the Jordan-Wigner transformation of the XY spin chain
\begin{align}
    H&= \sum_{m=1}^{n-1} \Big( \frac{1+\gamma}{2} \sigma_m^x \sigma_{m+1}^x +\frac{1-\gamma}{2} \sigma_m^y \sigma_{m+1}^y\Big) +\sum_{m=1}^{n}  h \sigma_m^z, 
\end{align}
with the Jordan-Wigner transformation given by:
\begin{align}
    \sigma_m^x=(-{\rm i})^{m-1} \prod_{j=1}^{2m-1}w_j, && \sigma_m^y=(-{\rm i})^{m-1} (\prod_{j=1}^{2m-2}w_j) w_{2m}.
\end{align}
%
Because of the non-locality of the Jordan-Wigner transformation, only linear driving on the first and the last spin can be included:
\begin{align}
         L_{\pm}^{1}&=\frac{1}{2} \sqrt{\Gamma_{\pm}^{1}} (w_{1} \mp {\rm i} w_2), && 
         L_{\pm}^{n}=\frac{1}{2} \sqrt{\Gamma_{\pm}^{n}} (w_{2n-1} \mp {\rm i} w_{2n})W . 
\end{align}
At the last site we encounter the parity operator $W=- (-{\rm i})^n w_1 w_2 \dots w_{2n}$, which however is unimportant since $L^{1,n}_\pm$ enter quadratically into the Lindblad equation, $ W W^\dagger=1$, and $W$ is a Casimir operator \cite{Prosen2008}.
%

We can use the above construction to treat the spin chain with open boundaries. 
Even though the XY spin ring can be exactly solved in a closed system context, this can no longer be done once we consider it as an an open quantum system.
Imposing periodic boundary conditions (PBC) $\sigma^\alpha_1=\sigma^\alpha_{n+1}$ results in 
$(-W)w_{2n+1}=w_1$, therefore the solution of the closed problem can be obtained from independent solutions of two fermionic problems: the odd (even) parity sector is the solution for the periodic(antiperiodic) fermionic problem \cite{Mazur1973,XYring}. 
The $L_\mu \rho L_\mu^\dagger$ term mixes the two parity sectors, and the driven spin problem with the periodic boundary condition becomes unsolvable in this manner.

\subsubsection{Correlations in the Open XY Spin Chain.}
The correlations containing $\sigma^z$ are easily expressible in terms of the the fermionic problem using $\sigma^z_m=- {\rm i} w_{2m-1}w_{2m}$, 
for example the $\sigma^z \sigma^z$ time-dependent correlation functions read
\begin{align}
   \langle \sigma^z_k(t) \sigma^z_j(0) \rangle=- \langle (w_{2k-1}w_{2k})(t)(w_{2j-1}w_{2j})(0)\rangle.
\end{align}
%

For comparison, the thermal correlation functions of the closed system in the thermodynamic limit are given by the following fermionic correlations \cite{Niemeijer1967, Mazur1973}:
\begin{align}
    \langle w_{2l-1}w_{2k} \rangle &= \frac{-{\rm i}}{2\pi} \int_{-\pi}^\pi \frac{(\cos \phi -h)+ {\rm i} \gamma \sin \phi}{\Lambda(\phi)}
    \tanh \left(\frac{\beta h \Lambda(\phi)}{2} \right)
    e^{{\rm i} (k-l) \phi} d \phi,
   \nonumber\\
    \Lambda(\phi) &= \sqrt{(\cos \phi -h^2)+  \gamma^2 \sin^2 \phi}.
\end{align}
 where $\Lambda(\phi)$ is the dispersion in the thermodynamic limit and furthermore $\langle w_{2l}w_{2k} \rangle = \langle w_{2l-1}w_{2k-1} \rangle = \delta_{l,k}$. 

Using the general framework presented in the previous section, we can numerically calculate the equal and \textit{non-equal} time correlation functions for systems larger than 1000 sites or look at the response to the local quench. Numerical results will be presented in the following section.

 \section{Numerical Results}
 \label{ch:results} 
 \subsection{Correlations in the NESS}
  \begin{figure}
    \centering
    \includegraphics[scale=.42]{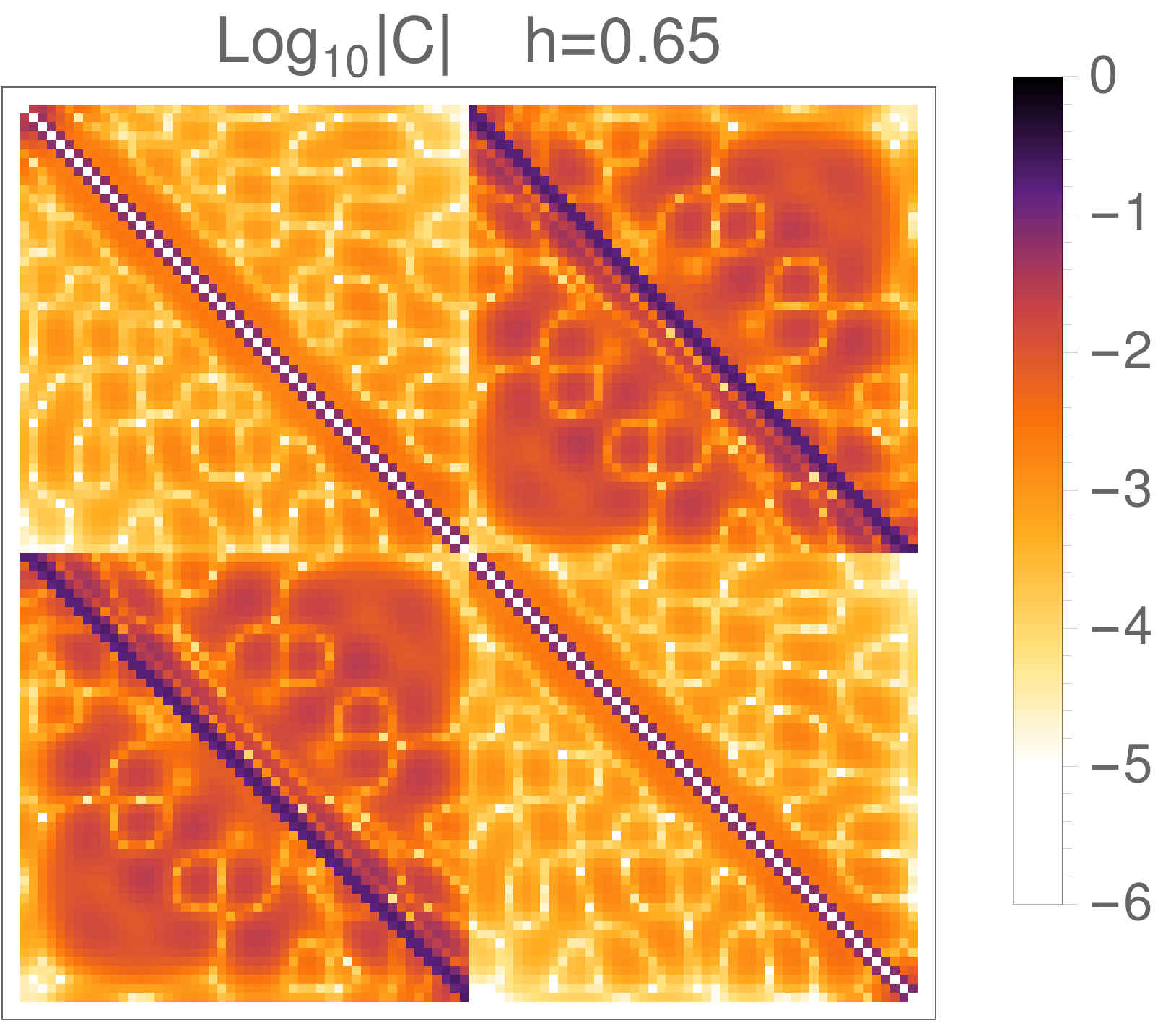}
    \includegraphics[scale=.42]{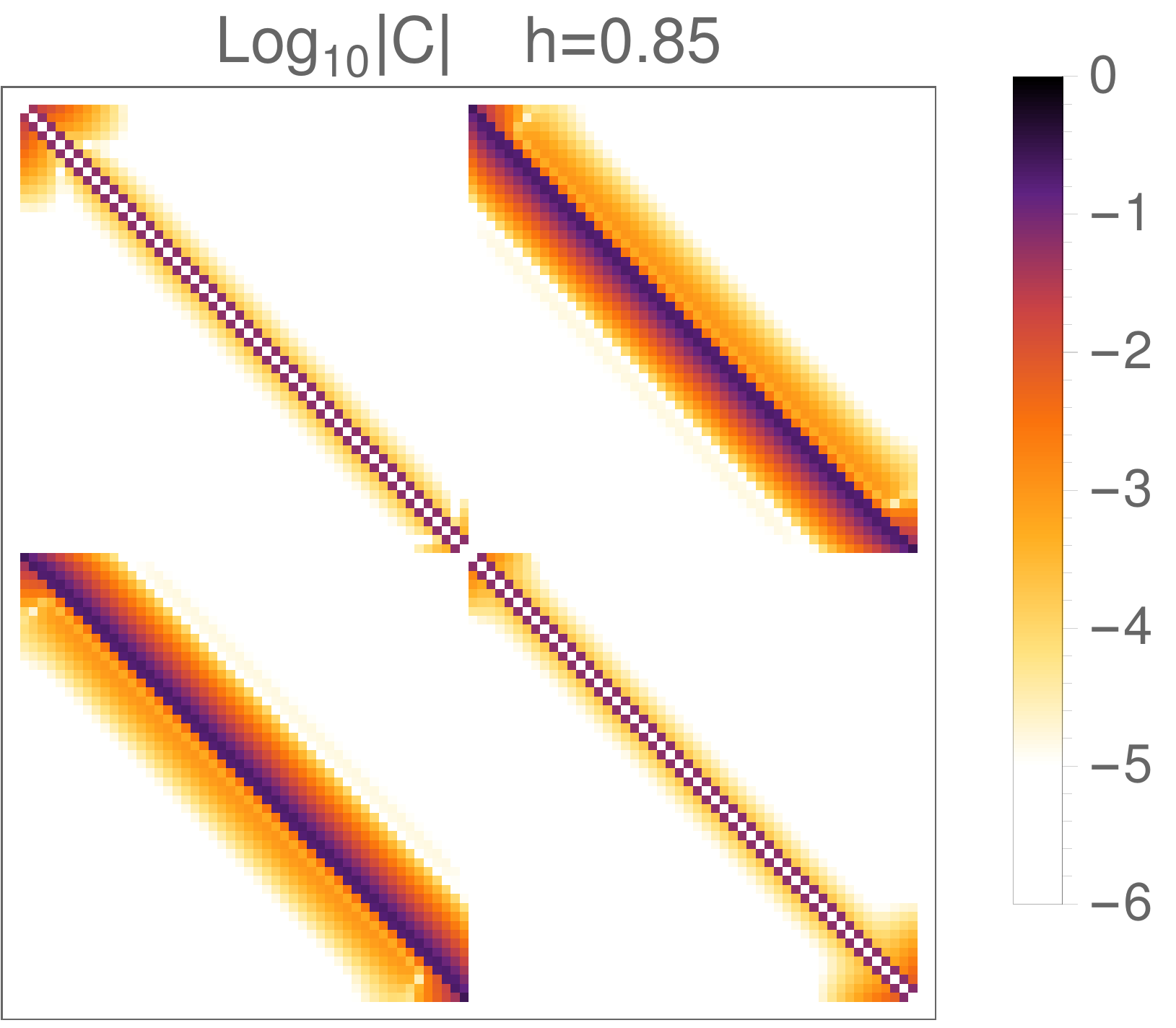}
\caption{The size of the equal-time two-point correlation function in the NESS (elements of the matrix $\textbf{C}^\infty$) for $h<h_c$ ({\it left}) and $h>h_c$ ({\it right}) of the open XY spin chain consisting of 50 spins with $\gamma=0.5$, $S_1=0.3$, $S_n=0.1$, $\beta_1=0.1$ and $\beta_n=5$. We have reordered the sequence of Majoranas as $\{w_1,w_3,\dots, w_{2n-1},w_2,w_4,\dots, w_{2n} \}$.
%
}
\label{fig:C2}
\end{figure}
 
\begin{figure}
    \centering
    \includegraphics[scale=.54]{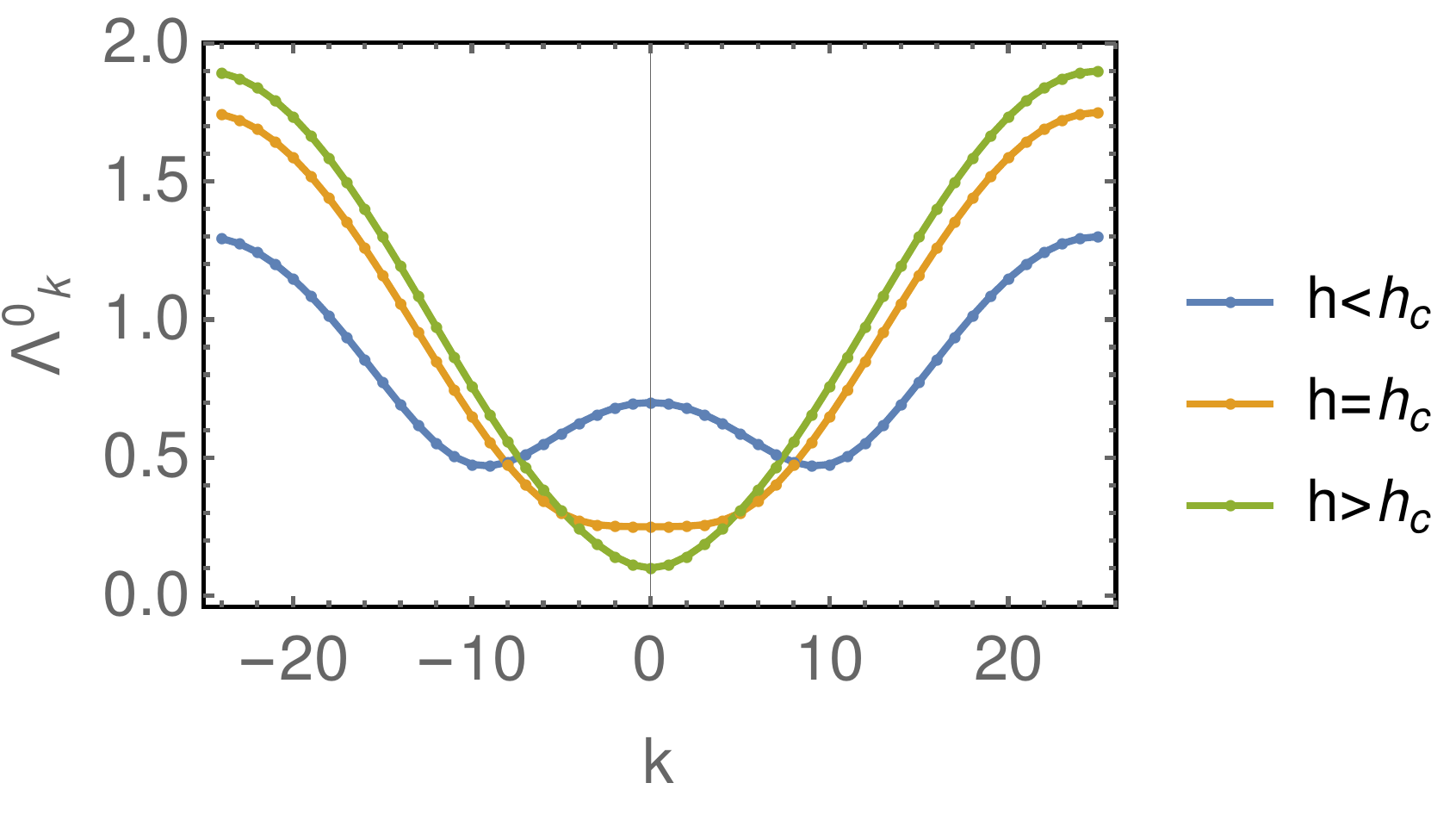}
    \caption{
    The free dispersion $\Lambda_k^0$ of the XY model that changes the number of local minima when we cross the non-equilibrium phase transition at $h_c=1-\gamma^2$.
    }
    \label{fig:dispersion}
\end{figure}

 As was argued in Refs.~\cite{Prosen2008a, PhysRevB.79.184416} the open XY spin chain exhibits a non-equilibrium phase transition at $h_c=1-\gamma^2$. The two point correlations are shown in the figure \ref{fig:C2}, and they are distinct in different phases. The long-range phase $h<h_c$ coincides with the appearance of the double minima in the dispersion relation as is indicated in the figure \ref{fig:dispersion}. For more details see \cite{Prosen2008a, PhysRevB.79.184416}.

 \subsection{Time-dependent Correlations} 
 
\begin{figure}
    \centering
    \includegraphics[scale=.38]{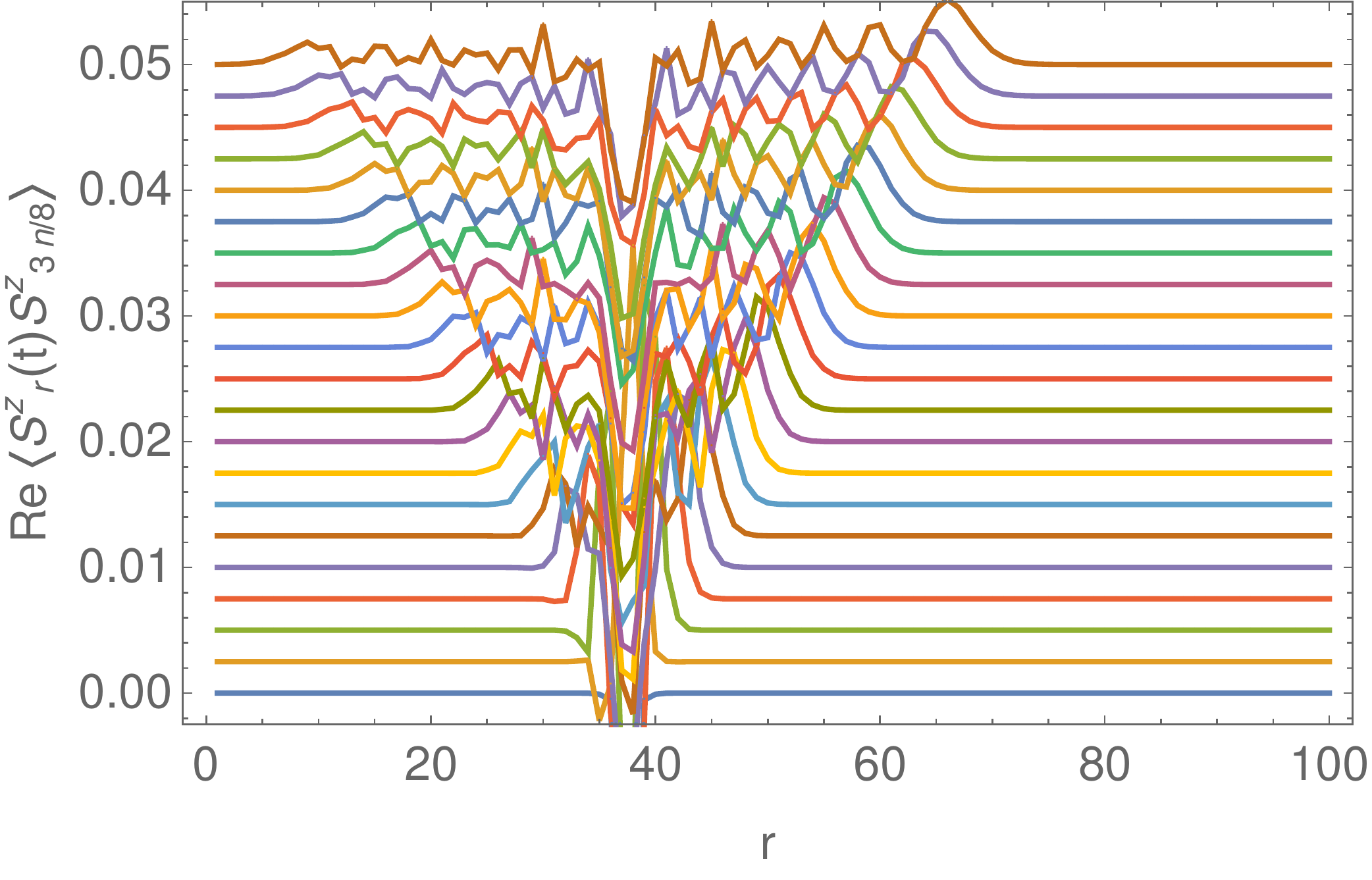}
    \includegraphics[scale=.38]{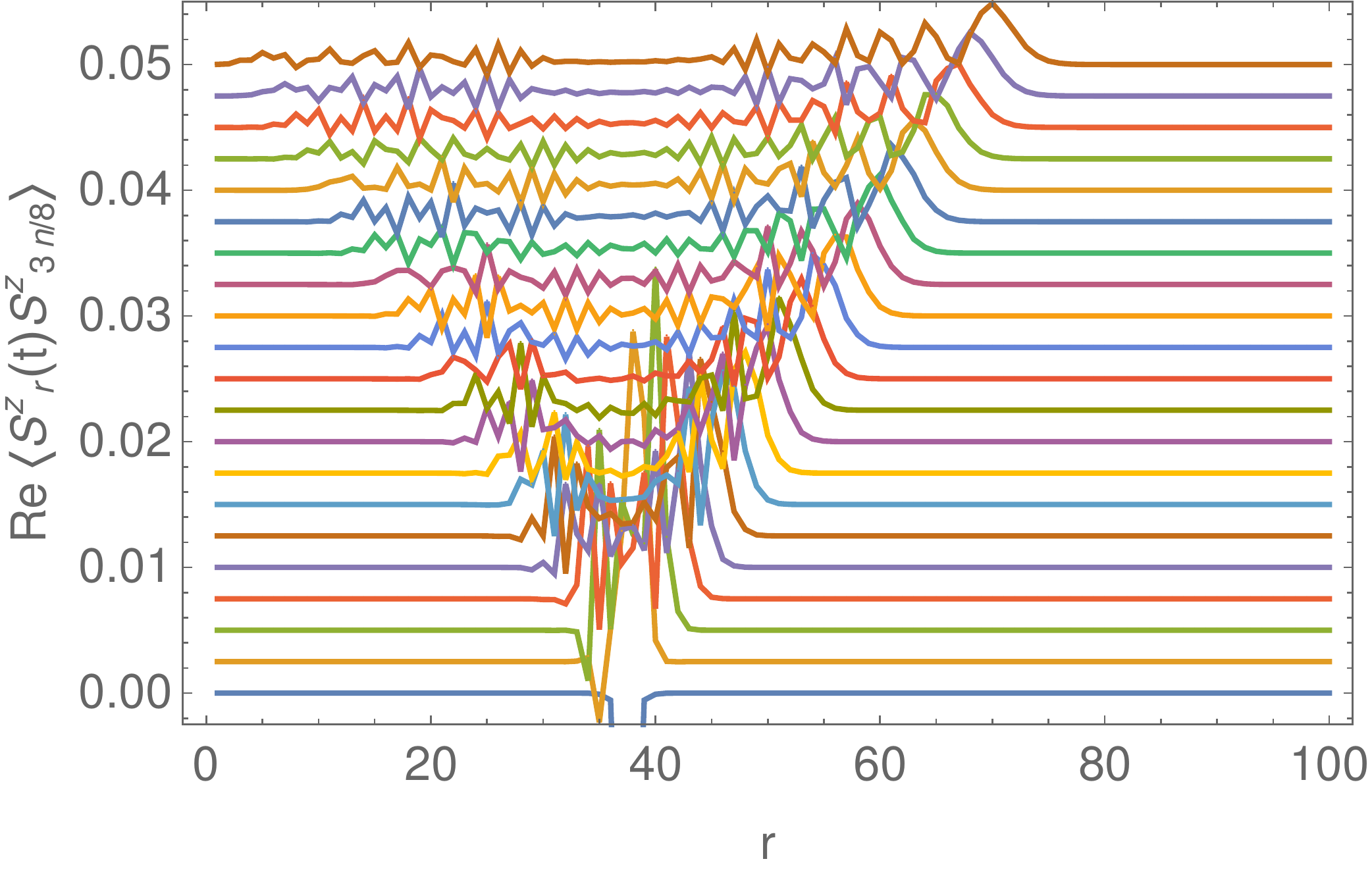}
\caption{A typical \textit{asymmetric} light cone responses of the real part of the connected time-dependent correlation function in the open XY spin chain for an asymmetric driving. Each new line is $\Delta t=1$ later and is lifted for better visibility. 
In the {\it left} ({\it right}) figure  $h=0.6$ ($h=0.9$), so we are in different non-equilibrium phases ($h_c=0.75$) and the central region behaves differently. Parameters are $\gamma=0.5$, $n=100$, $S_1=S_n=1$, $\beta_{1}=0.5$ and $\beta_{n}=2$.
}
\label{fig:LC}
\end{figure}

\begin{figure}
\centering
    \includegraphics[scale=.40]{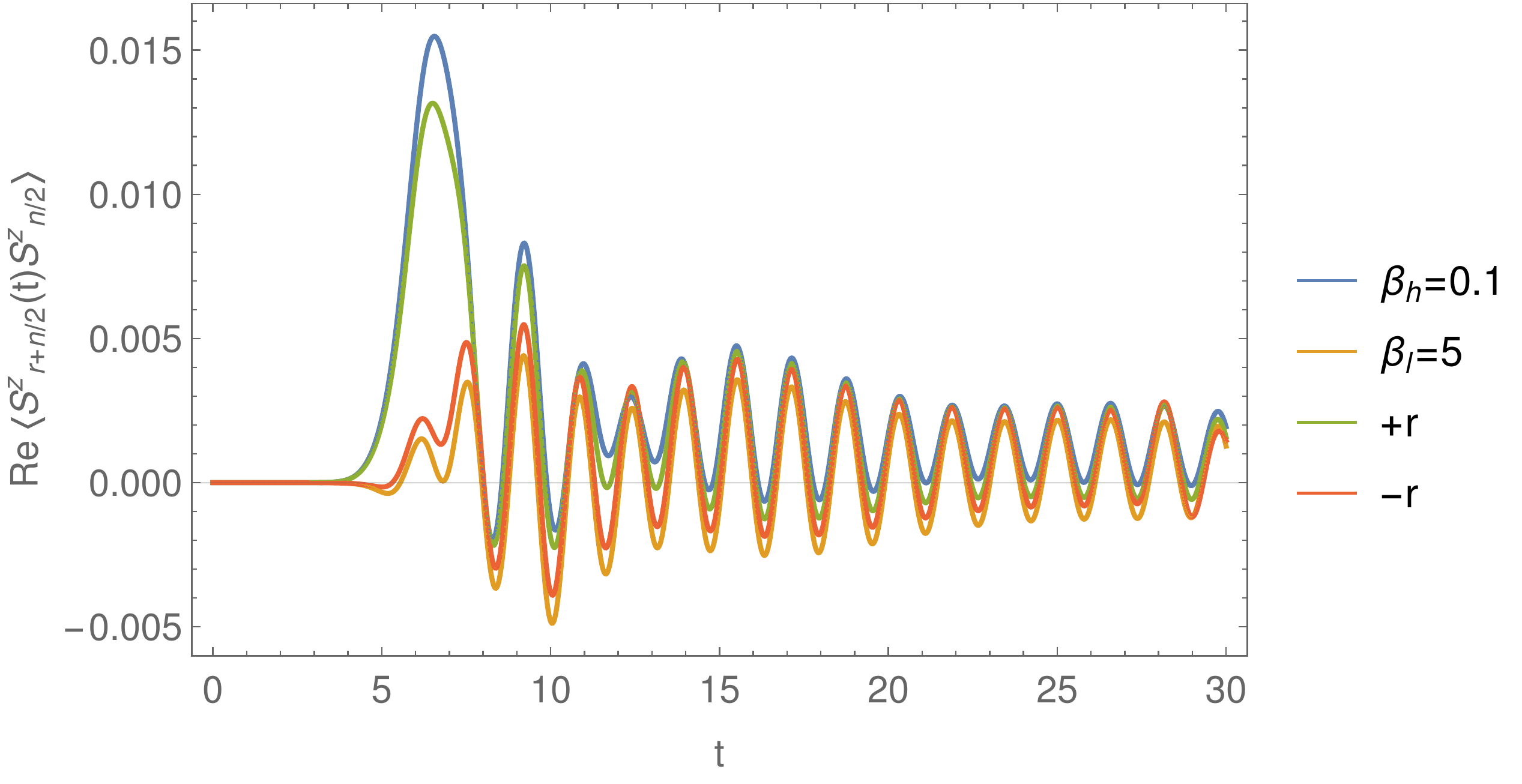}
    \includegraphics[scale=.40]{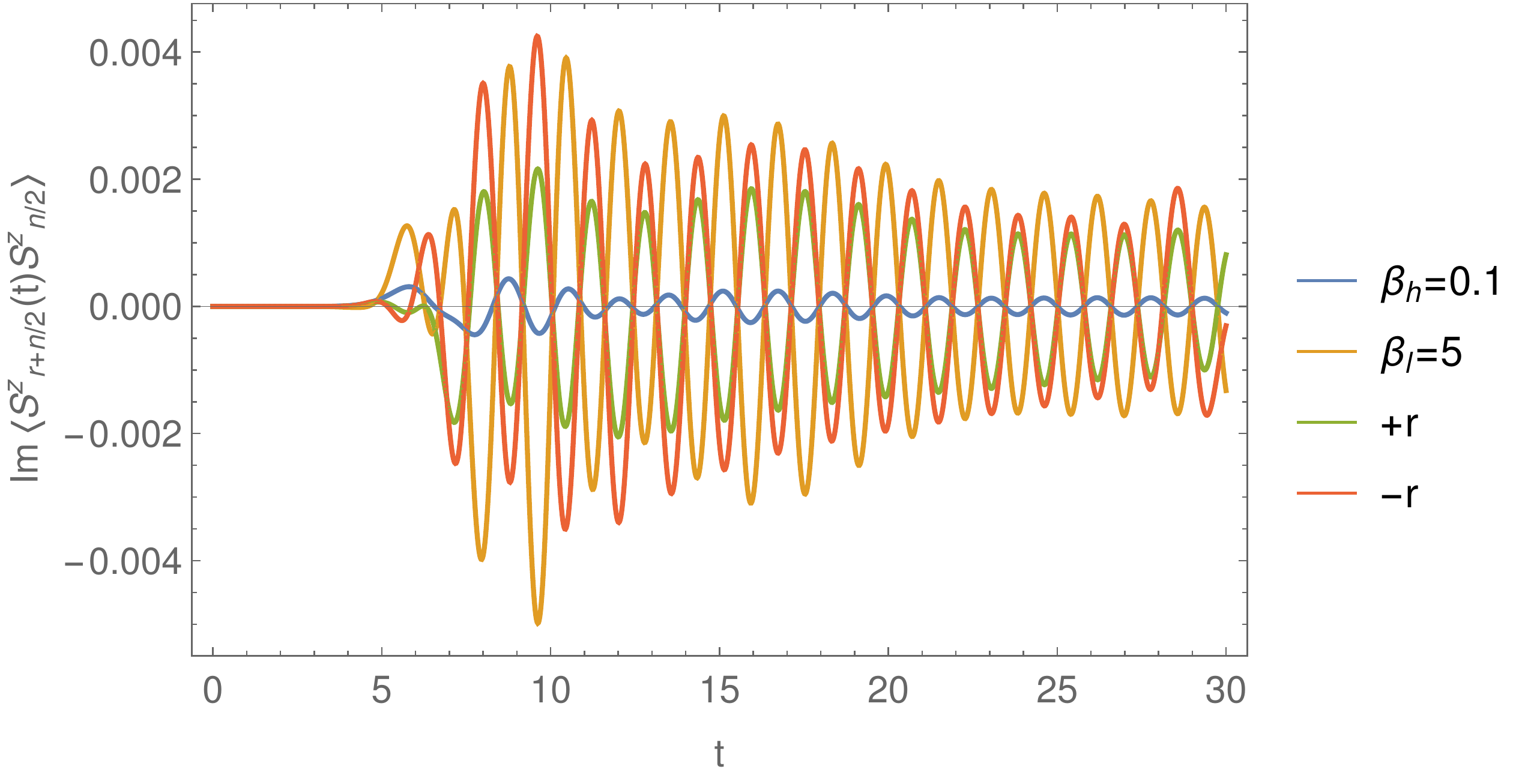}
\caption{
The real and imaginary part of the connected time-dependent correlation functions of the open XY spin chain for the baths at different temperatures. We see that $\langle S_{r+n/2}^z(t) S_{n/2}^z \rangle$ labeled $+r$ and $\langle S_{-r+n/2}^z(t) S_{n/2}^z \rangle$ labeled $-r$ are different. 
The real part of the correlation to the right $+r$ ($-r$) is very similar to the thermal response of the system with $\beta_h$ ($\beta_l$), corresponding to the temperature of left bath $\beta_1=0.1$ (right bath $\beta_n=5$). We can imagine the bath sending thermal quasi-particles to the right (left). 
Other parameters are $r=10$, $n=64$, $\gamma=0.5$, $h=0.9$, $S_1=0.3$ and $S_n=1$.
}
\label{fig:response}
\end{figure}

In the figures \ref{fig:LC} and \ref{fig:response} we show the time-dependent correlation function (also interpreted as a local quench) of the open XY model, which shows the light cone behaviour. The response to the left and to the right has the same velocity, but the response is asymmetric for the asymmetric driving as can be clearly seen in the figure \ref{fig:response}.

\subsection{Open XY model with the Dzyaloshinskii-Moriya(DM) Interactions} 

An interesting solvable extension ot the XY model is the addition of the Dzyaloshinskii-Moriya (DM) interactions that break the space reflection symmetry.

We focus on the special form $\vec{D}= D \hat{e}_z$:
\begin{align}
\begin{split}
    H_{DM}&= D \sum_m^n \hat{e}_z (\vec{\sigma}_m \times \vec{\sigma}_{m+1})
    =D \sum_m^n (\sigma_m^x \sigma_{m+1}^y - \sigma_m^y \sigma_{m+1}^x)\\
    &=- {\rm i} D \sum_l^{2n} w_{l} w_{l+2} 
\end{split}
\end{align}

The model is exactly solvable in the thermodynamic limit of a closed system, where the dispersion is given by \cite{DM1}:
\begin{align}
    \Lambda(\phi)=  \left| 2 D \sin \phi + \sqrt{(h-\cos \phi)^2+\gamma^2 \sin^2 \phi} \right|.
\end{align}

\begin{figure}
    \centering
    \includegraphics[scale=0.7]{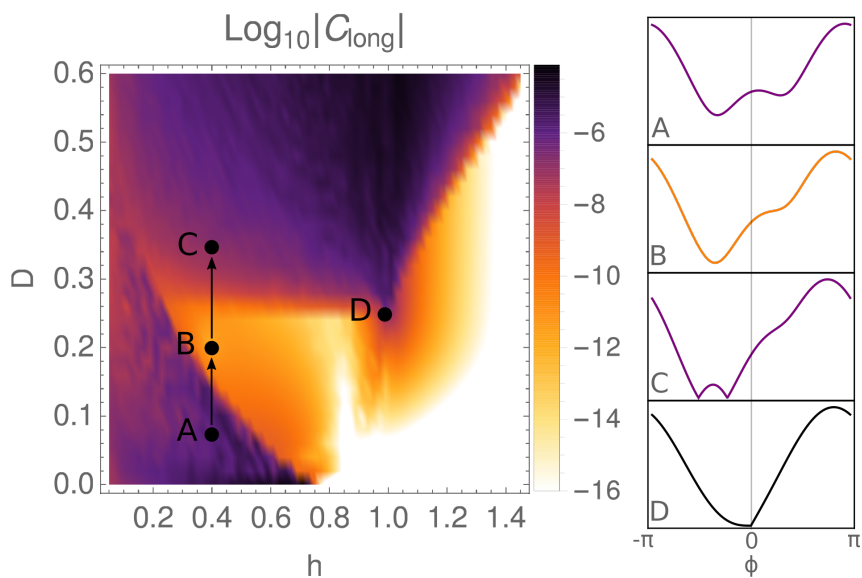}

    \caption{The phase diagram of the open XY model with the Dzyaloshinskii-Moriya interactions at fixed anisotropy $\gamma=0.5$. The average correlations between the middle site and the last quarter of the spin chain with 100 spins are plotted. The darker region with the residual long-range correlations coincides with the appearance of double minima in the dispersion relation which is plotted schematically on the right.
    When we go from a point A to B, at the phase boundary the second minimum of the dispersion relation disappears.
     When we go from B to C, the left minimum hits the zero-level and bifurcates to a pair of Dirac points. D is a special point, and the boundary line continuing through it is the transition of the B $\to$ C type.
Parameters are
    $n=100$, $\gamma=0.5$, $S_1=S_n=1$, $\beta_1=\beta_n=2$, $h=1.05$ and $D=0.3$.
}
    \label{fig:DM1}
\end{figure}

In figure \ref{fig:DM1} we show the static (equal-time) correlations in the NESS, which were used as the indicator of the non-equilibrium phase transition in the open XY model \cite{Prosen2008a}. We notice a novel non-equilibrium phase transition between the two NESS phases with different long-range correlations, similar to the known transition from the XY model without a DM term, that is still present. Interestingly, the new phase transition again coincides with the emergence of two local minima in the free model's dispersion (this time from the absolute value), as is illustrated in the figure \ref{fig:DM1} on the right.
There we get two Dirac points at nontrivial values ($\neq 0,\pi$) of $\phi$, which indicated different low energy behaviour of the model.

\begin{figure}
\centering
    \includegraphics[scale=0.4]{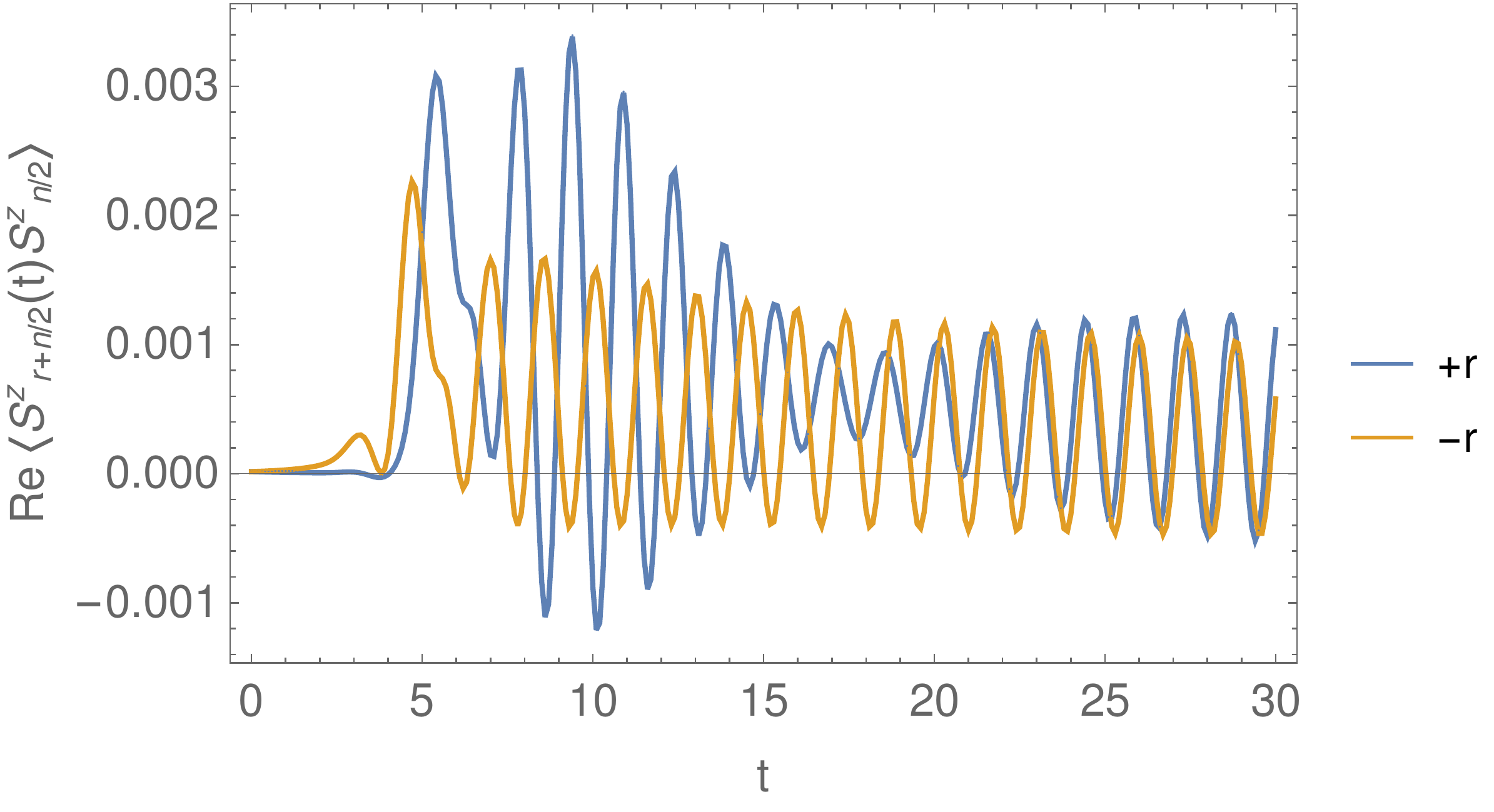}
\caption{
    Even though we have a symmetric driving the dynamic response is asymmetric, because the DM interactions break reflection symmetry. Parameters are as in the figure \ref{fig:DM1} and $r=10$.
    }
    \label{fig:DM2}
\end{figure}

It is important that the reflection symmetry is broken by the DM interactions, namely space reflection changes the sign of the coupling constant $D$. Therefore, we get an asymmetric dynamic response even for a symmetric driving, as is shown in the figure \ref{fig:DM2}.



\subsection{Two-dimensoinal Model} 
\begin{figure}
    \centering
    \includegraphics[scale=.3]{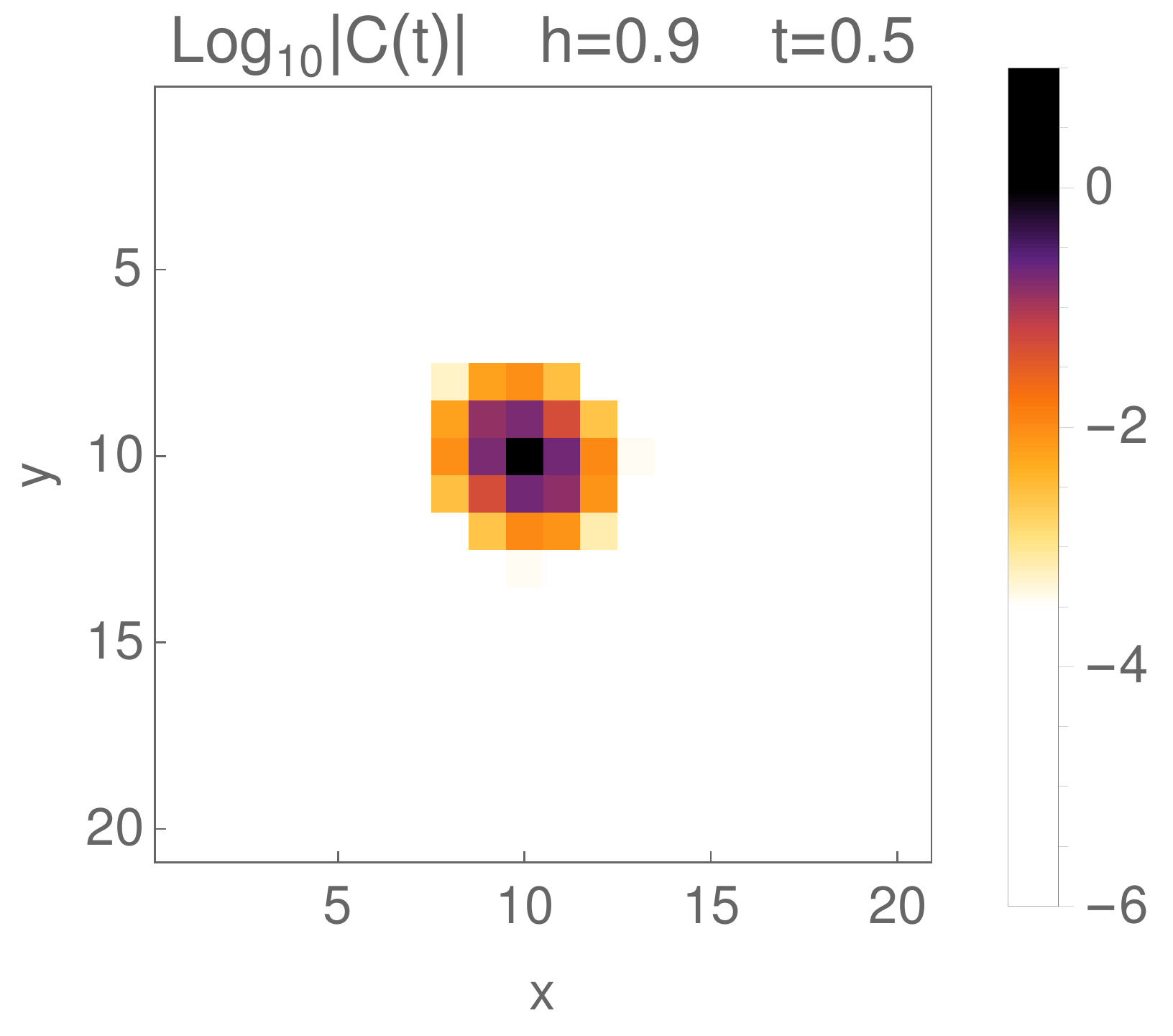}
    \includegraphics[scale=.3]{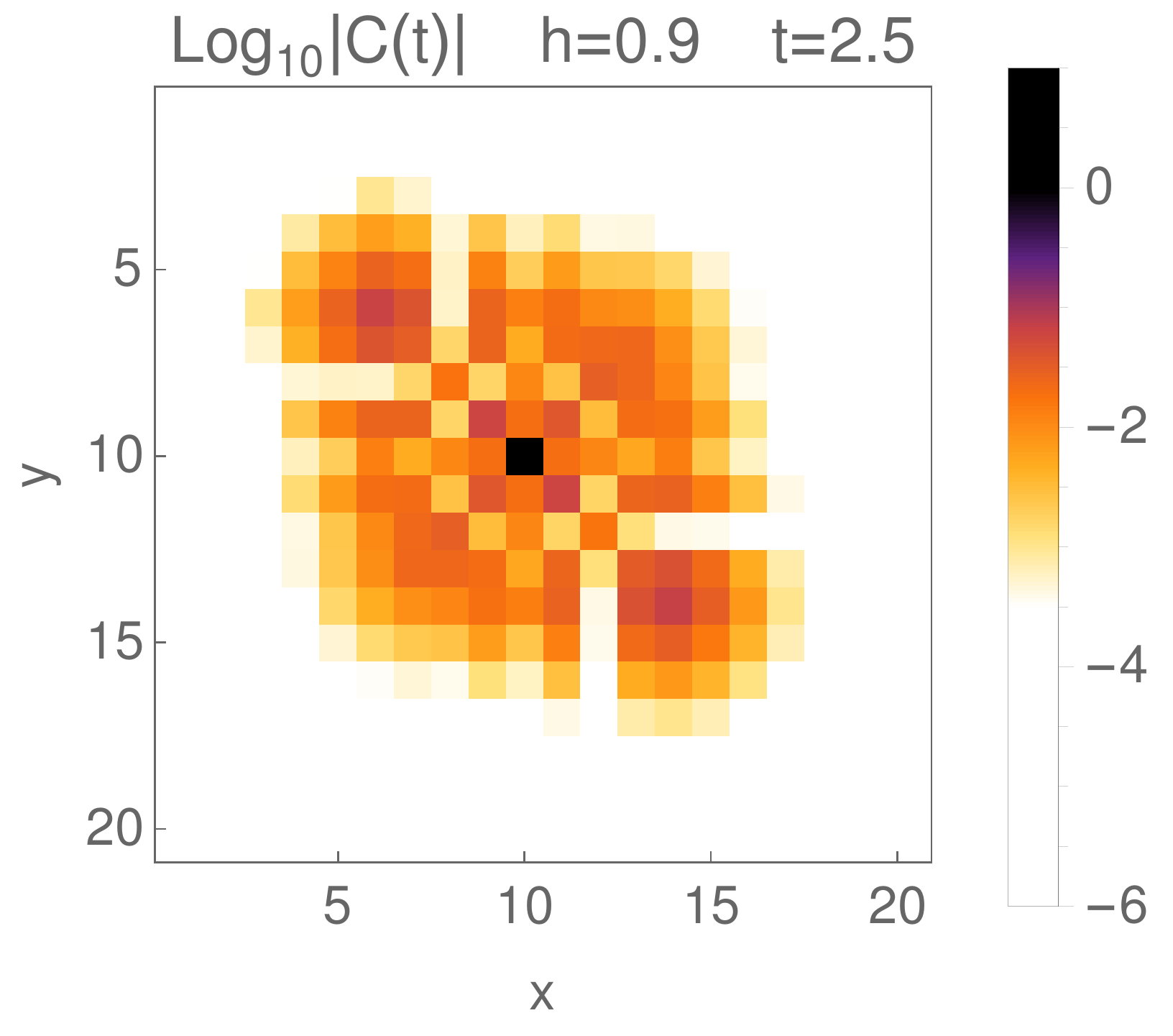}
    \includegraphics[scale=.3]{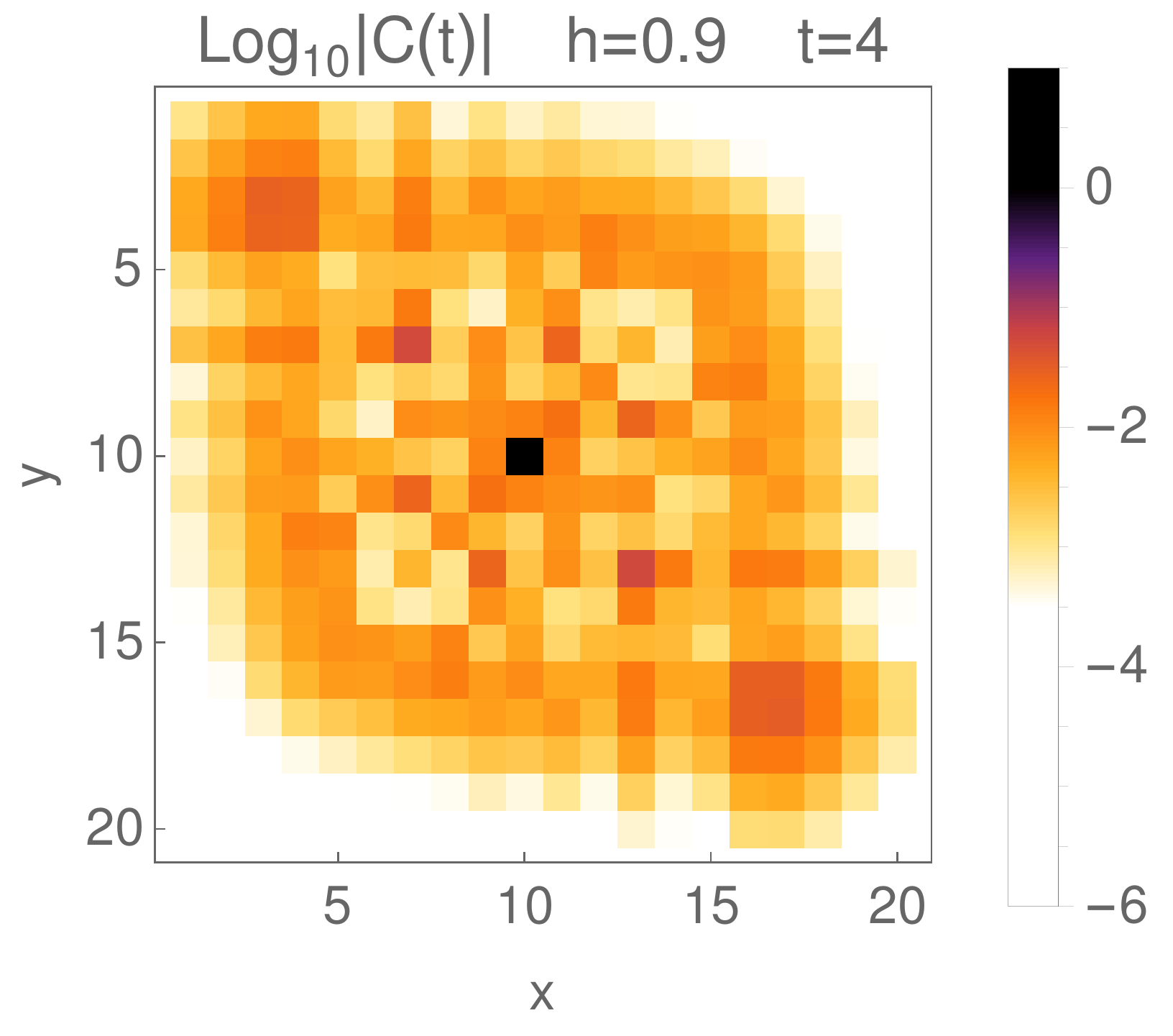}
    \includegraphics[scale=.36]{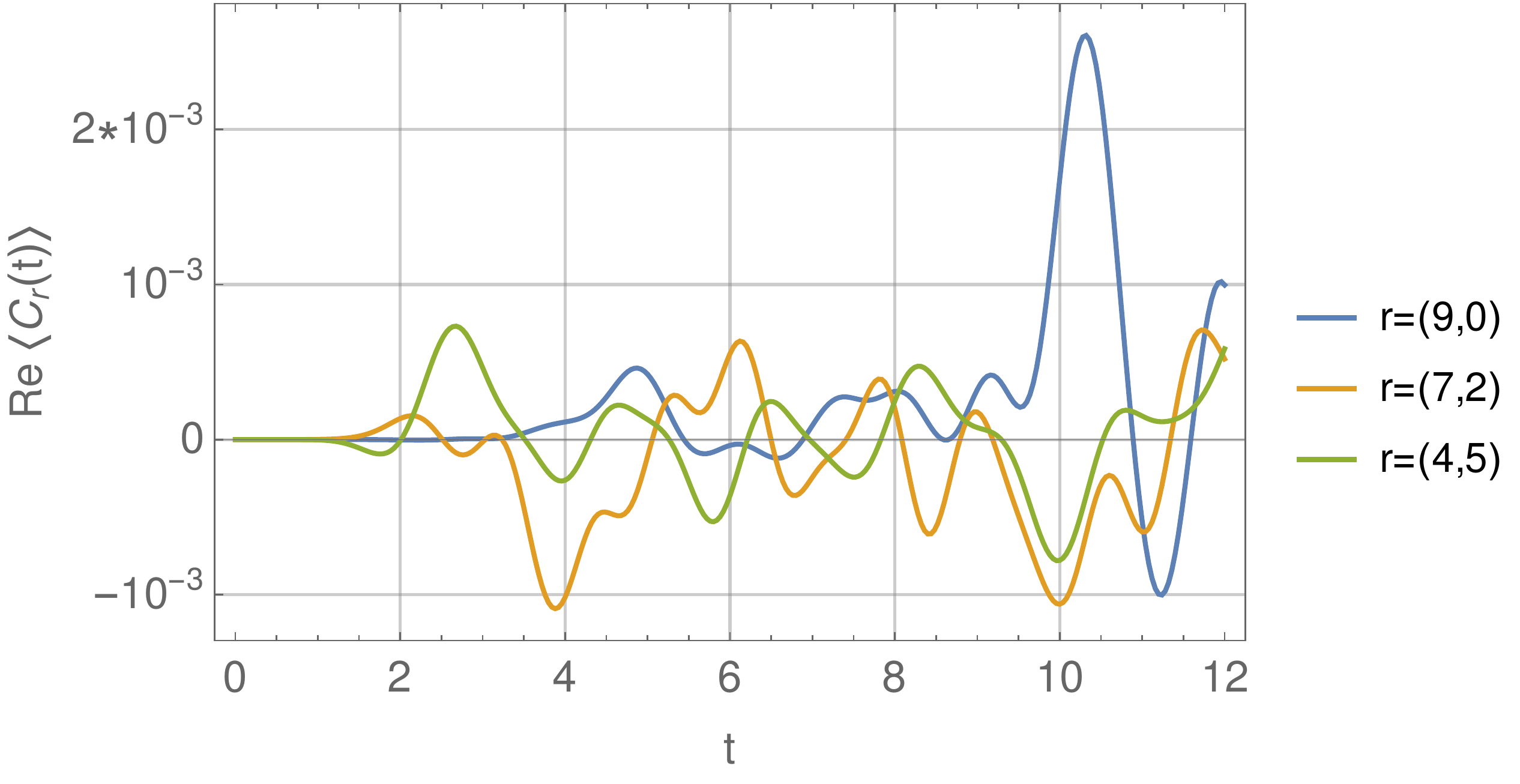}
    \includegraphics[scale=.36]{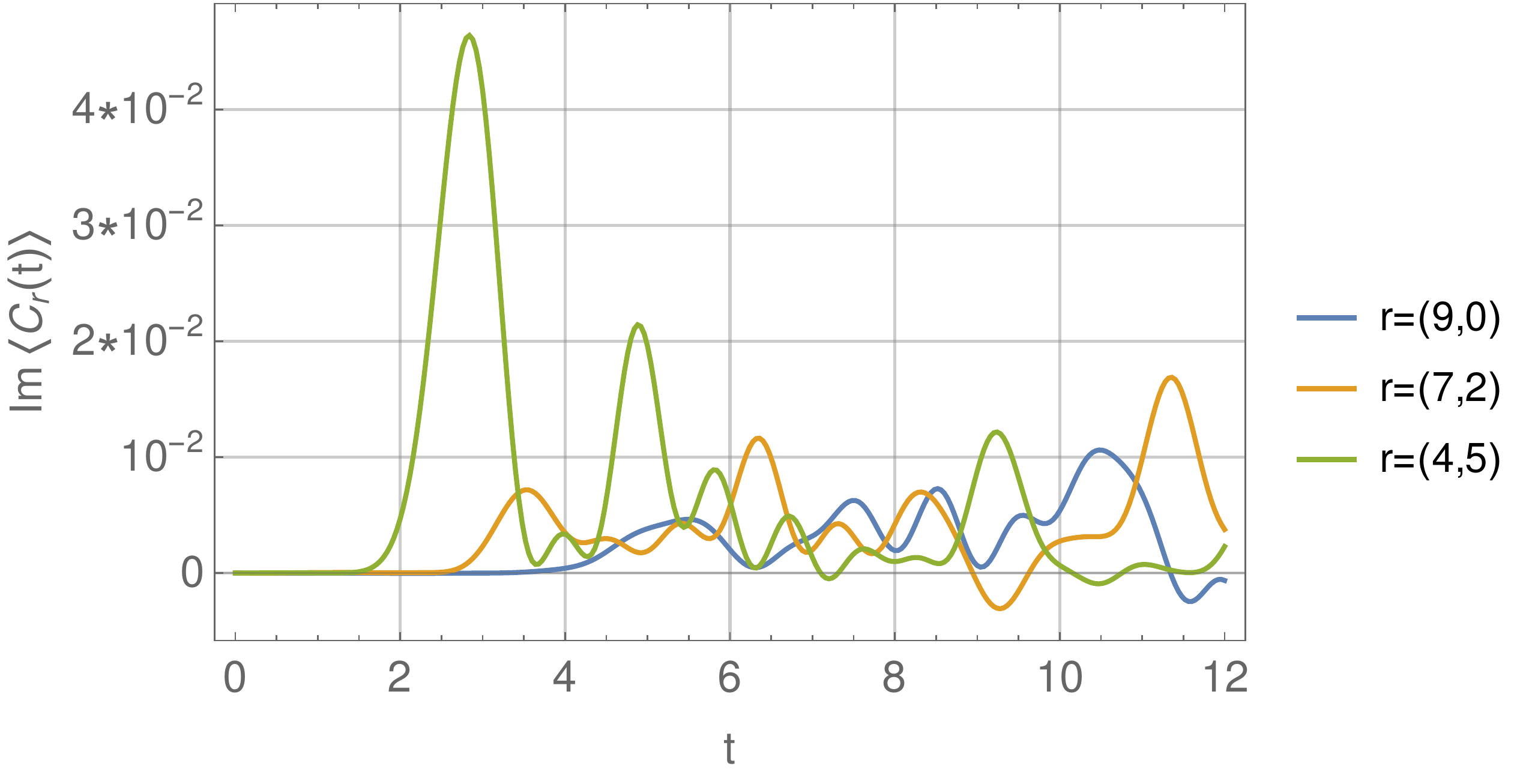}
\caption{The dynamic response of a two-dimensional model given by the Hamiltonian \eqref{eq:Hami2D} and the dissipation in the top-left and the bottom-right corners. We plot the connected time-dependent correlation between the $(x,y)$ at the time $t$ and $(\frac{n}{2},\frac{n}{2})$ at the time $0$ given by $C(t)^c$ defined in the equation \eqref{eq:defCct}.
We notice a non-uniform $2+1$ dimensional light cone, since the correlations spread faster in the directions of the driving.
The {\it bottom} two pictures show real and imaginary parts of the dynamic response at the points $(\frac{n}{2},\frac{n}{2})+r$ for different vectors $r$, which show different responses.
Parameters are $n = 20$, $\gamma=0.5$, $S_{(1,1)}=0.3$, $S_{(n,n)}=0.1$, $\beta_{(1,1)}=0.1$ and $\beta_{(n,n)}=5$.
%
}
\label{fig:2D}
\end{figure}

Next, we present the time-dependent correlation function (figure \ref{fig:2D}) for a two-dimensional problem given by the Hamiltonian 
\begin{align}
     H= &\sum_{\langle j,k \rangle} (c_j^\dagger c_{k}+ \gamma c_j^\dagger c_{k}^\dagger + h. c. ) + 2 h \sum_j c_j^\dagger c_j,
     \label{eq:Hami2D}
\end{align}
with $\langle j,k \rangle$ denoting the nearest neighbours on a square $n \times n$ lattice $\{1,2\ldots n\}\times\{1,2\ldots n\}$ with open boundary conditions. The system is driven at the top-left $(1,1)$ and the bottom-right $(n,n)$ corners with the Lindblad operators given by the equation \eqref{eq:driving}. We looked at the connected part of the density-density correlations at the different times
\begin{align}
  C(t)^c_{(x,y)}=\frac{\langle (w_{2r-1} w_{2r})(t) (w_{2s-1} w_{2s})(0) \rangle -\langle (w_{2r-1} w_{2r})(0) (w_{2s-1} w_{2s})(0) \rangle}{ \langle w_{2s-1} w_{2s} \rangle}
  \label{eq:defCct}
\end{align}
where the indices of the Majoranas are determined by $r= x+ n y$ and $s= \frac{n}{2}+ n \frac{n}{2}$.

We notice that the spreading of the correlations is non-uniform, since they spread faster in the directions of the driving.

\clearpage
\section{Analytic Solutions for the Weakly Coupled Fermionic Ring}
\label{sec:weak XY}

Analytical results provide us with important new information and insights. Here we are interested in systems governed by the fermionic Hamiltonian \eqref{eq:Hami}, which is connected to the XY spin problem, when there is no dissipation.
There is no know analytical solution for the non-driven (closed) XY spin chain with the open boundary conditions (except for $h=0$ case solved by Fagotti \cite{Fagotti2016}), so we have treated this problem only numerically. Note that obtaining the spectrum analytically is possible for the XX spin chain ($\gamma=0$) for special values of driving parameters $S_1$ and $S_n$ which was done by Guo and Poletti \cite{Guo2017}. 

Our approach in this section is to use the analytical solution for the XY problem in the closed quantum system \cite{Lieb1961, Mazur1973, Niemeijer1967} to analytically diagonalise the matrix $\textbf{X}$, Eq. (\ref{Xmatrix}), and then include the coupling with baths as a small perturbation, therefore solving the weakly coupled fermionic problem. Since the driving couples different parity sectors, our result does not apply to the driven XY spin ring. 
For simplicity we will assume that the number of sites $n$ is even.

\subsection{Spectrum of the Matrix $\textbf{X}$}
It is convenient to reorder the sequence of Majoranas as $\{w_1,w_3,\dots, w_{2n-1},w_2,w_4,\dots, w_{2n} \}= \{w_{(1,1)}, \dots, w_{(n,1)}, w_{(1,2)},\dots, w_{(n,2)}  \}$ 
so the reordered matrix $\textbf{X}=-2 {\rm i} \textbf{H} +2\textbf{M}_r$ reads
\begin{align}
\textbf{X}&=\begin{pmatrix} \textbf{D}& \textbf{T}\\-\textbf{T}^T&\textbf{D}\end{pmatrix}, &&
 \textbf{D}=\frac{1}{4} \text{diag} \left( S_1, S_2, \dots, S_n \right),
 \end{align}
 \begin{align}
    \textbf{T}=
    \begin{pmatrix}
    -h& \frac{1-\gamma}{2}&0&   \dots& 0&\frac{1+\gamma}{2}\\
    \frac{1+\gamma}{2}&-h&\frac{1-\gamma}{2}&0 &\dots&0\\
    0&\frac{1+\gamma}{2}&-h&\frac{1-\gamma}{2}&0 &\dots
    \\
     \ddots&\ddots&\ddots&\ddots&\ddots&\ddots\\
     0&     \dots &0& \frac{1+\gamma}{2}&-h&\frac{1-\gamma}{2}\\
     \frac{1-\gamma}{2} &0    &\dots & 0& \frac{1+\gamma}{2}&-h
    \end{pmatrix}.
\end{align}
The matrix $\textbf{D}$ encodes the dissipation, and the circulant $n \times n$ matrix $\textbf{T}$ (tri-diagonal Toeplitz matrix for open BC) encodes the Hamiltonian.
For a weakly coupled fermionic ring, we can treat the coupling to the baths as a perturbation and calculate the real corrections to the imaginary eigenvalues, which govern the decay to the steady state.
A closed system corresponding to $\textbf{X}_0$ with $\textbf{D}=0$ case can be diagonalised as \cite{Niemeijer1967}:
\begin{align}
    \textbf{P} \textbf{X}_0 \textbf{P}^\dagger= \begin{pmatrix} {\rm i} \Lambda^0  &0          \\
                                     0          & -{\rm i} \Lambda^0        \end{pmatrix}, 
                                     && 
   \textbf{P}=\begin{pmatrix}\Phi&{\rm i} \Psi\\
    \Phi&-{\rm i} \Psi\end{pmatrix}, 
    && 
    \textbf{P} \textbf{P}^\dagger=\textbf{1}_{2n},
\end{align}
$\Psi \Psi^T=\Phi \Phi^T=\textbf{1}_{n}$ and $\Phi \textbf{T}= \Lambda^0 \Psi$. The eigenvalues and eigenvectors are:
\begin{align}
    (\Lambda)_{k,k}^0&=\Lambda_k^0= \sqrt{(\cos \phi_k -h^2)+  \gamma^2 \sin^2 \phi_k }, &
    \lambda_k&=\frac{1}{2} \arctan \frac{\gamma \sin \phi_k}{\cos \phi_k-h},\\
    \Psi_{kj}&=\sqrt{\frac{2}{n}} \cos \left( j \phi_k- \frac{\pi}{4}-\lambda_k \right), &
    \Phi_{kj}&=\sqrt{\frac{2}{n}} \cos \left( j \phi_k- \frac{\pi}{4}+\lambda_k \right).
    \label{eq:def psi phi}
\end{align}
The parameter $\lambda_k$ captures the nonzero anisotropy $\gamma$, with a function $\arctan$ defined such that $0<\lambda_k \leq \pi$.
The quantization condition gives $\phi_k=k\frac{2\pi}{n}$ for the periodic boundary conditions of the fermionic problem ($\phi_k=(k-\frac{1}{2})\frac{2\pi}{n}$ for the anti-periodic boundary conditions); $k=1,\dots,n$.

We will now compute the first order corrections for the weakly coupled XY ring using the first order perturbation theory.
%
%
The eigenvectors in the non-degenerate subspaces remain the same, so an attempt to diagonalise $\textbf{X}$ with unperturbed eigenvectors results in:
\begin{align}
   \textbf{P} \textbf{X} \textbf{P}^\dagger= 
    \begin{pmatrix} 
    {\rm i} \Lambda^0+
    \frac{\Phi \textbf{D} \Phi^T+\Psi \textbf{D} \Psi^T}{2}&
    \frac{\Phi \textbf{D} \Phi^T-\Psi \textbf{D} \Psi^T}{2}\\
    \frac{\Phi \textbf{D} \Phi^T-\Psi \textbf{D} \Psi^T}{2}&
     -{\rm i} \Lambda^0+\frac{\Phi \textbf{D} \Phi^T+\Psi \textbf{D} \Psi^T}{2}
     \end{pmatrix}.
\end{align}
In order to correctly calculate the first order corrections, we need to take into account that the spectrum of the closed system is degenerate:
$\Lambda_k^0= \Lambda_{n-k}^0$ (for aPBC $\Lambda_k^0= \Lambda_{n-k+1}^0$).
Let us denote $\textbf{K}=\frac{\Phi \textbf{D} \Phi^T+\Psi \textbf{D} \Psi^T}{2}$ which can be calculated as
\begin{align}
    \textbf{K}_{k,l}&= \sum_{j=1}^n  \frac{S_j}{8} \left( \Psi_{k,j}  \Psi_{l,j}+\Phi_{k,j}  \Phi_{l,j} \right)\\
                    &=\sum_{j=1}^n \frac{S_j}{4n}  \left\{ \cos (j(\phi_k-\phi_l)) \cos(\lambda_k-\lambda_l)+ \sin (j (\phi_k+\phi_l)) \cos (\lambda_k+\lambda_l) \right\}. \nonumber
\end{align}
Elements of the degenerate subspaces can be further simplified using $\lambda_{n-k}=-\lambda_k+\pi$:
\begin{align}
    \textbf{K}_{k,k}&=\frac{1}{4n} \sum_{j=1}^n S_j \left(1+ \sin  2 j \phi_k \cos2 \lambda_k \right)
    =\frac{1}{4n}\left( \sum_{j=1}^n S_j + (\vec{S}_k)_y \cos2\lambda_k \right), 
    \nonumber\\
     \textbf{K}_{n-k,k}&= \textbf{K}_{k,n-k}= 
     -\frac{\cos2 \lambda_k}{4n}  \sum_{j=1}^n S_j \cos  2 j \phi_k = -\frac{\cos2 \lambda_k}{4n} (\vec{S}_k)_x,
     \\
     \vec{S}_k&=\begin{pmatrix}\sum_{j=1}^n S_j \cos 2 j \phi_k \\ \sum_{j=1}^n S_j \sin 2 j \phi_k\end{pmatrix}=\begin{pmatrix}S_x\\S_y\end{pmatrix}.
\end{align}
In the last equalities we introduced a convenient vector notation with the dependence on $k$ implicit, which we use to express the first order real corrections to eigenvalues and eigenvectors of the degenerate subspaces as:
\begin{align}
     \Lambda^{1}_{\pm,k} &= \sum_j\frac{S_j}{2n}  \pm \frac{|\vec{S}|}{2n} \cos 2\lambda_k ,
    && 
    v_{\pm,k}=\frac{1}{\sqrt{|v_{\pm,k}|}} \begin{pmatrix}S_y \pm |\vec{S}|\\-S_x\end{pmatrix}.
\end{align}
Therefore the correct eigenvectors that diagonalise $\textbf{X}_0$ and the correction $\textbf{K}$ in the degenerate subspaces are:
\begin{align}
        \Phi_{k,l}^\pm&=v_{\pm,k,1} \Phi_{k,l} +v_{\pm,k,2} \Phi_{n-k,l}, &
        \Psi_{k,l}^\pm=v_{\pm,k,1} \Psi_{k,l} +v_{\pm,k,2} \Psi_{n-k,l}.
        \label{eq:phi_pm}
\end{align}
The new basis is labeled by the index pair $\{\pm,k \}$; $k=1,\dots, \frac{n}{2}$ instead of one index $k=1,\dots, n$.

\subsubsection{Special Cases.}
We will have a closer look at the three special cases:
\begin{itemize}
\item[a)]  $S_n=S\neq 0$
\item[b)]  $S_n=S_1 = S \neq 0$
\item[c)]   $S_j=S$ for every $j$
\end{itemize}
\begin{figure}
    \centering
    \includegraphics[width=.66\textwidth]{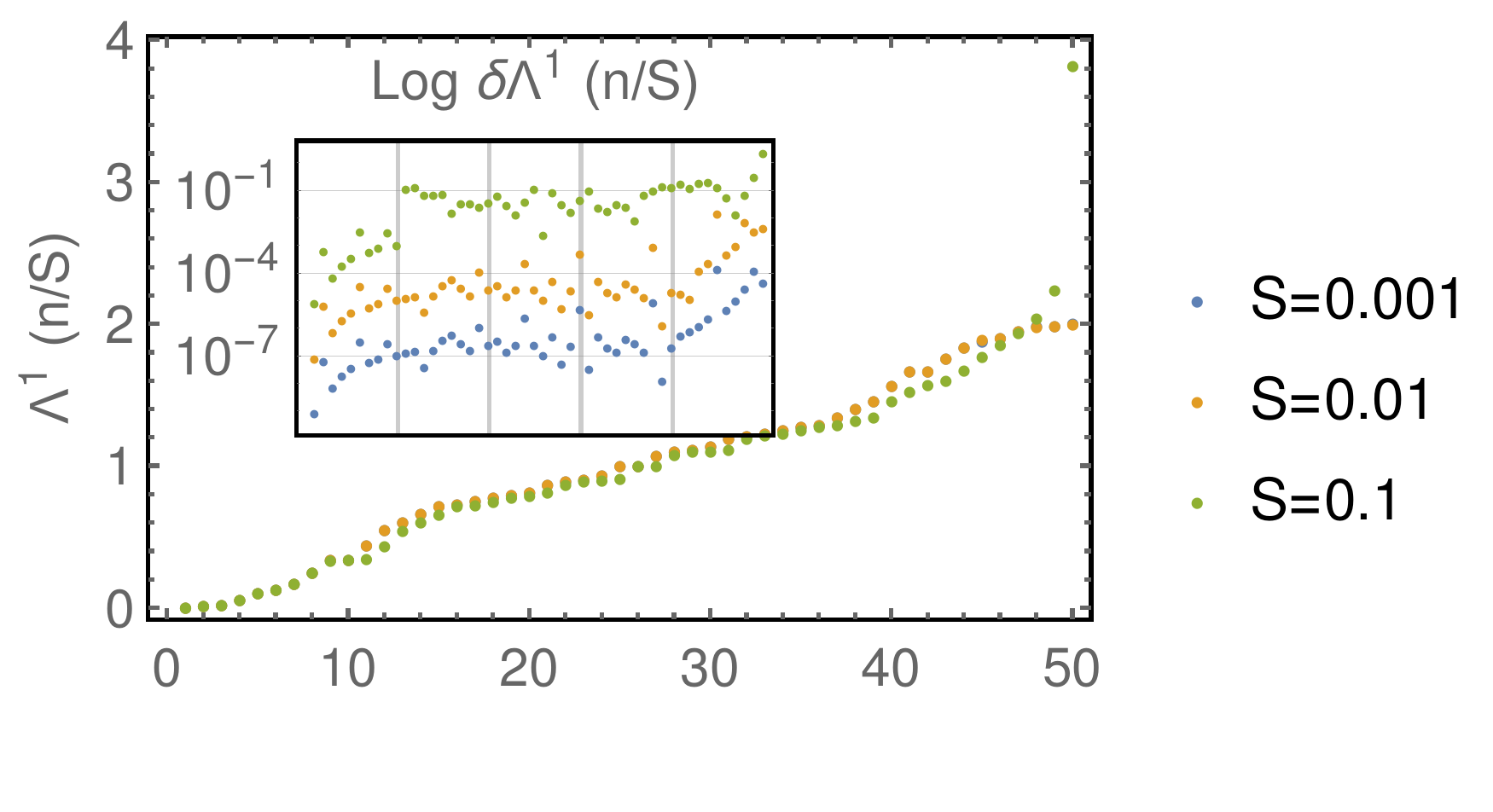}
    \includegraphics[width=.66\textwidth]{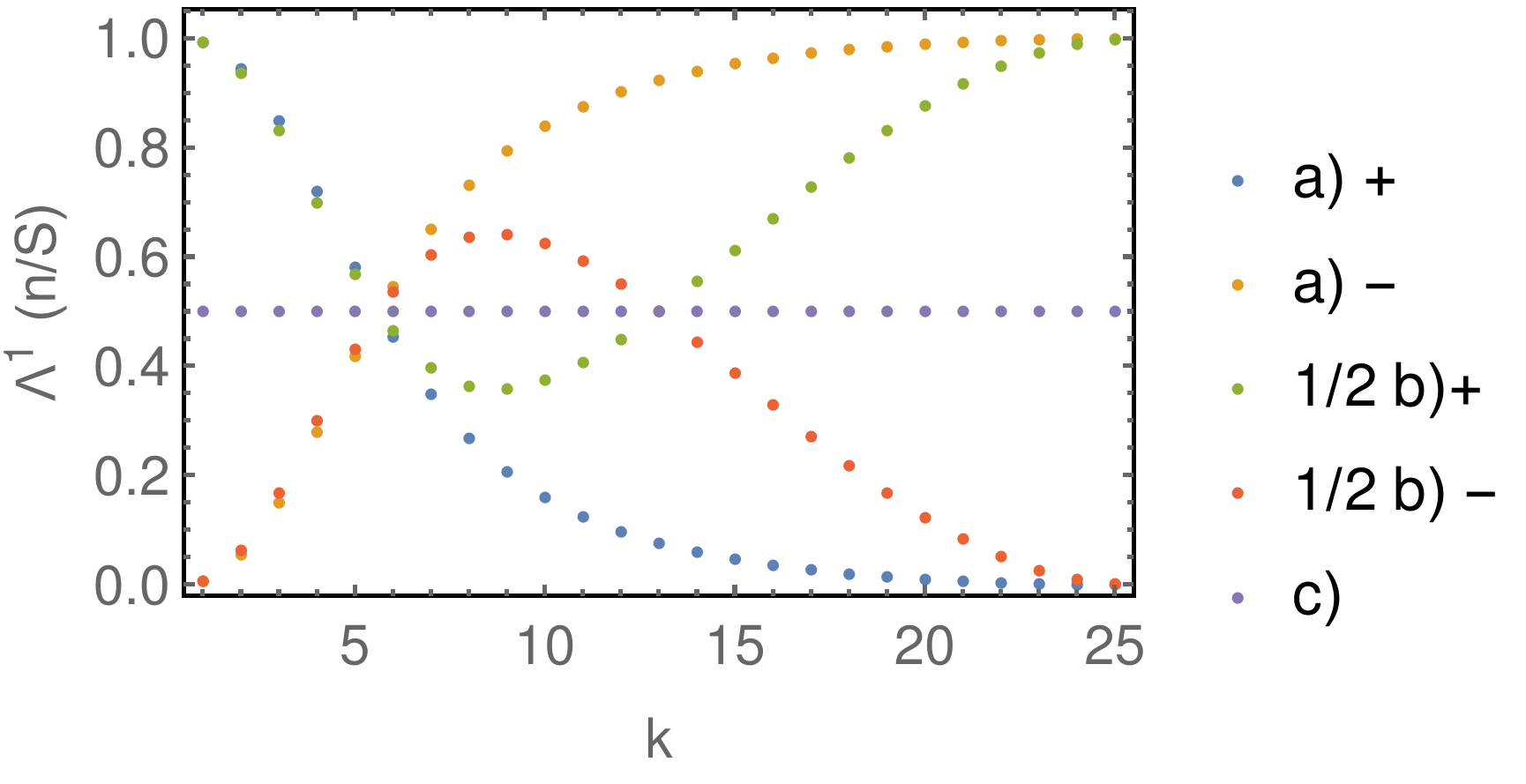}
    \caption{{\it Top}: numerically calculated first order corrections $\Lambda^1$ resized by a factor $\frac{n}{S}$ and ordered by the size for different magnitudes of the coupling $S$ in the case b). The shape stays mostly the same, with the exception of the largest correction, which gets much larger for larger values of $S$. In the inset we show the logarithm of the error of analytical values given by \eqref{eq:L1 case b} for small coupling S (coupling at sites $1$ and $n$, $h=.8$, $\gamma=0.5$, $n=50$).
    {\it Bottom}: the analytical corrections in the cases a) (rescaled by $\frac{n}{S}$), b) (rescaled by $\frac{n}{2S}$) and c) (rescaled only by $S$).}
    \label{fig:ev}
\end{figure}
with all other (unspecified) coupling constants $S_j$ equal to zero. 

In the case a) $\Lambda^{1}_{\pm,k} =\frac{S}{2n} \left(1 \pm \cos 2\lambda_k \right)$ so  $\Lambda^{1}_{+,k} =\frac{S}{n} \cos^2 \lambda_k$, 
$\Lambda^{1}_{-,k} =\frac{S}{n} \sin^2 \lambda_k$ and eigenvectors are $(1,1)^T$ and $(1,-1)^T$ for $k=1, \dots, n/2$.

On the other hand, in the case b) we have:
\begin{align}
\Lambda^{1}_{\pm,k}  &=\frac{S}{n} \left(1 \mp \cos 2\lambda_k \cos \phi_k \right),&&  
    \vec{S}=2 S \cos \phi_k \begin{pmatrix}\cos \phi_k\\\sin \phi_k \end{pmatrix}, 
    \label{eq:L1 case b}\\
    v_{\pm,k}&=\frac{1}{\sqrt{2(1\pm \sin \phi_k)}}\begin{pmatrix}\sin \phi_k \pm 1\\- \cos \phi_k\end{pmatrix}.
\end{align}

Case c) results in $\vec{S}=0$ and $\Lambda^{1}= \frac{S}{2}$ with no dependence on $n$. The results are illustrated in figure \ref{fig:ev}.

\subsection{Static Correlation Functions in the NESS}
As seen in the section \ref{sec:Correlation functions} (and noting in our case $\tilde{\textbf{P}}=\textbf{P}^\dagger$), the correlation functions are given by the matrix $\textbf{C}$ (we will skip the superscript $^\infty$), which is determined by:
\begin{align}
    ( \textbf{P} \textbf{C} \textbf{P}^\dagger)_{jk}&=4 {\rm i} \frac{(\textbf{P} \textbf{M}_i \textbf{P}^\dagger)_{jk}}{\Lambda_j+\Lambda_k^*},
    &
\textbf{P}&=\frac{1}{\sqrt{2}}\begin{pmatrix} 
&\Phi^+      &   &{\rm i}\Psi^+  &   \\ 
&\Phi^-    &     &{\rm i}\Psi^- &    \\ 
&\Phi^+    &     &-{\rm i}\Psi^+   &  \\ 
&\Phi^-    &     &-{\rm i}\Psi^-    & \\ 
\end{pmatrix}
,
\label{eq:PCP}
\end{align}
%
%
with $\Phi^\pm,\Psi^\pm$ defined in \eqref{eq:phi_pm}. 
To calculate the correlation functions we need to compute the relevant elements of $(\textbf{P} \textbf{M}_i \textbf{P}^\dagger)$. 
Here
$\textbf{M}_i=\begin{pmatrix}0   & \tilde{M_i}\\ -\tilde{M}_i &0 \end{pmatrix}$ with $\tilde{M_i}=\text{diag}(\dots, \frac{S_j}{4} \tanh \beta_j h,\dots)$.
Namely,
\begin{align}
    \textbf{P} \textbf{M}_i \textbf{P}^\dagger=\frac{1}{2} \begin{pmatrix} 
    -{\rm i} (\Phi \tilde{M}_i \Psi^\dagger+\Psi \tilde{M}_i \Phi^\dagger)&  {\rm i}  (\Phi \tilde{M}_i \Psi^\dagger-\Psi \tilde{M}_i \Phi^\dagger)\\
    -{\rm i} (\Phi \tilde{M}_i \Psi^\dagger-\Psi \tilde{M}_i \Phi^\dagger)  &  {\rm i}(\Phi \tilde{M}_i \Psi^\dagger+\Psi \tilde{M}_i \Phi^\dagger)
    \end{pmatrix},
    \label{eq:PMiP}
    \end{align}

where we skipped indices $\pm$ for better visibility and wrote $\Phi= \begin{pmatrix}\Phi^+ \\ \Phi^-\end{pmatrix}$.
The largest contribution for $\textbf{P} \textbf{C} \textbf{P}^\dagger$ comes from $\Lambda_j+\Lambda_k^* \approx 0 $. Up to the first order correction in $S$ it means that ${\rm i} (\Lambda_j^0-\Lambda_k^0)+\Lambda_j^1+\Lambda_k^1 \approx S $ and we get the largest correction in the degenerate subspaces. 
Using the explicit expressions
\begin{align}
\begin{split}
    \Phi_{kj}^\pm&=\frac{1}{\sqrt{n}|v_{\pm,k}|} \Big\{ 
    \left(v_{\pm,k,1}+v_{\pm,k,2} \right)(\cos \lambda_k \sin j\phi_k-\sin\lambda_k \cos j \phi_k)
    \\
    &+\left(v_{\pm,k,1}-v_{\pm,k,2} \right)( \cos \lambda_k \cos j\phi_k+\sin\lambda_k \sin j \phi_k)
    \Big\},
    \end{split}
\end{align}
and the same expression for $\Psi$ with $-\lambda_k$ instead of $\lambda_k$ we derive:
\begin{align}
\begin{split}
    \Phi_{kj}^\pm \Psi_{kj}^\pm &= \frac{1}{n} \left\{ 
    \cos2\lambda_k\pm \frac{S_x}{|\vec{S}|} \cos2j \phi_k \pm \frac{S_y}{|\vec{S}|} \sin2j\phi_k    \right\}, 
    \\
    \Phi_{kj}^\pm \Psi_{kj}^\mp &= \frac{1}{n} \left\{\mp \frac{S_x}{|S_x|} \sin 2 \lambda_k- \frac{S_x}{|\vec{S}|} \sin 2j\phi_k
       \right\},
       \end{split}
\end{align}
which is all that we need since $\tilde{M}_i$ is a diagonal matrix and we are interested in the first nontrivial order in $S$. 

\subsubsection{Static Correlation Functions: Case a).} 
\begin{figure}
    \centering
    \includegraphics[trim={0 -20 0 0},clip,scale=0.38]{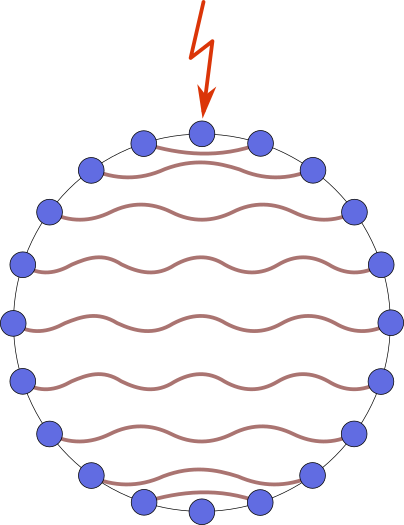} \qquad
    \includegraphics[trim={0 35 0 0},clip,scale=0.45]{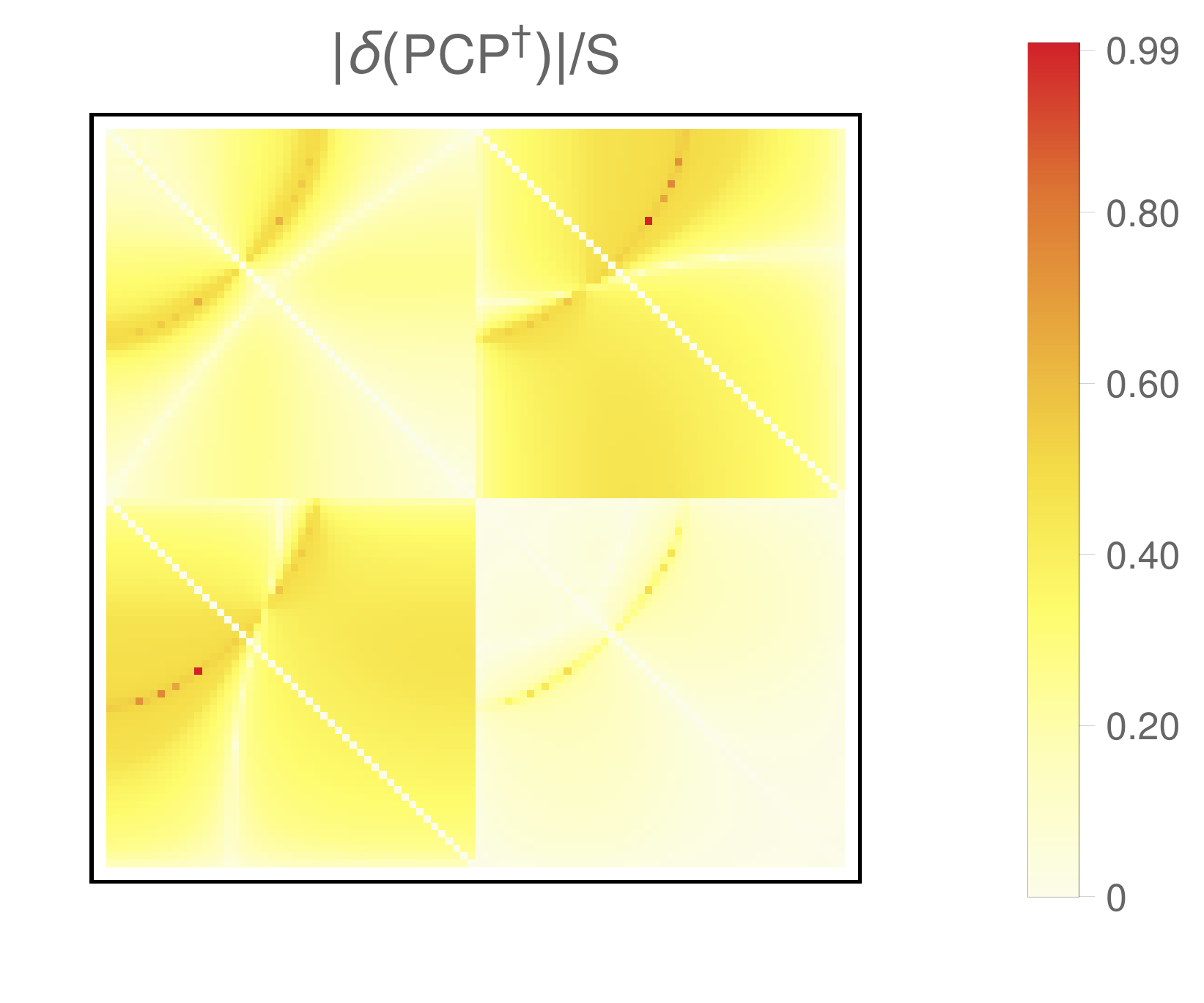}
    \caption{{\it Left}: a graphics showing the interesting correlations in the case a), which are between even-odd Majoranas and the same in size. Arrow shows the driven site.
    {\it Right}: the second order corrections in the case b) for $h<h_c$ are enhanced along the arcs which are indicating long distance correlations in the NESS.}
    \label{fig:cartoon}
\end{figure}

In the case a) of dissipation at a single site the equation \eqref{eq:PCP} results in:
\begin{align}
    (\textbf{P} \textbf{C} \textbf{P}^\dagger)_{k,k}^{\pm,\pm}&=\pm 4 \tanh h \beta ,
\end{align}
with other elements equal to 0 and $k=1,\dots, n/2$ (for $k>n$ we get additional minus sign as seen in Eq.~\eqref{eq:PMiP}). Interestingly, other elements are exactly zero and the result is robust against stronger coupling (the numerics and analytics suggests that the result is correct and exact for any coupling). The matrix $\textbf{C}$ encoding the correlations in the NESS is simple:
\begin{align}
    \textbf{C}_{(k,1),(l,2)}&=-{\rm i} \ {\tanh h \beta} \left( \delta_{k,n-l
    }- \delta_{k,n} \delta_{l,n} \right),
\end{align}
and all other nonzero elements are determined by the asymmetry of the matrix $\textbf{C}$ ($\textbf{C}^T=-\textbf{C}$). We denoted by 1 (2) odd (even) Majorana subspaces ($w_{(j,1)}=w_{2j-1}$ and $w_{(j,2)}=w_{2j}$).
This gives very interesting long-range correlations between even-odd Majoranas that are at the same difference from the site $n$ in the opposite directions as shown in figure \ref{fig:cartoon} and does not depend on the bath coupling strength $S$.

\subsubsection{Static Correlation Functions: Case b).}
The first order results for the special case b) ($S_1=S_n=S$) are given by:
\begin{align}
    (\textbf{P} \textbf{C} \textbf{P}^\dagger)_{k,k}^{+,+}&=\frac{ (\tanh h\beta_1+\tanh h\beta_n) (\cos 2 \lambda_k + \cos \phi_k ) }{2 \left(1 + \cos 2\lambda_k \cos \phi_k \right)},
    \\
    (\textbf{P} \textbf{C} \textbf{P}^\dagger)_{k,k}^{-,-}&=\frac{ (\tanh h\beta_1+\tanh h\beta_n) (\cos 2 \lambda_k - \cos \phi_k ) }{2 \left(1 - \cos 2\lambda_k \cos \phi_k \right)},
    \\
    (\textbf{P} \textbf{C} \textbf{P}^\dagger)_{k,k}^{+,-}&=\frac{(\tanh h\beta_1-\tanh h\beta_n) (\sin \phi_k ) }{2} \frac{\cos \phi_k}{|\cos \phi_k|}.
\end{align}
In this case the resulting expression is more complicated:
\begin{align}
  \textbf{C}_{(k,1), (l,2)}&= 
  {-{\rm i}}\frac{\tanh h \beta_1-\tanh h \beta_n}{4} \left( \delta_{k+l,2+n} - \delta_{k+l,n} \right)
  +
   {\rm i} \frac{\tanh h \beta_1+\tanh h \beta_n}{2n} \nonumber \\
   &\sum_{a=1}^{n/2}
   \frac{1}{(\cos \phi_a-h)^2+\gamma^2}  
   \Big( 2 (\cos \phi_a-h)^2 
   \left[ \cos (k{-l})\phi_a +\frac{\gamma \sin \phi_a \sin (k{-l})\phi_a}{\cos \phi_a-h}
   \right]
    \nonumber\\
    & \qquad \qquad \qquad \qquad \qquad +\gamma^2 (\cos (k+l)\phi_a+\cos (k+l+2)\phi_a)
   \Big).
\end{align}
There are again interesting correlations between the sites $j$ and $n-j$ (and $n-j+2$), similar to the case a). This is the first order result, and the higher orders are nontrivial. In fact it misses an important feature: change of the long distance correlation behaviour when we cross the non-equilibrium phase transition. This is a $\mathcal{O}(S)$ effect in the weakly coupled regime (it does not depend on $S$ for larger couplings or the open chain). 
It is a consequence of the form of the dispersion for $h<h_c$ (see figure \ref{fig:dispersion}) where we have an additional almost degeneracy $\Lambda_k=\Lambda_l$ when $\cos \phi_l+ \cos \phi_k=2\frac{h}{h_c}$ and we get large contributions for the elements $(\textbf{P} \textbf{C} \textbf{P}^\dagger)_{k,l}$ that are the reason for the correlations at the larger distances. They are shown in figure \ref{fig:cartoon} on the right and are arc shaped.

\subsection{Validity of the First Order Perturbation Theory}
 Here we will have a closer look at the validity of the perturbation theory, with a focus on the large $n$ (thermodynamic) limit. This limit can be problematic, because the unperturbed energy level spacing gets smaller and smaller, and from some point on, we may need to treat the neighbouring energy levels as almost degenerate and use the degenerate perturbation theory.
 
 In general we can say that the first order perturbation theory will work if the first order correction for the eigenvectors are small:
 \begin{align}
   (\vec{\Psi}^\pm_k)^1&=\sum_{\substack{l \neq k \\ \sigma= \pm} }\frac{\textbf{K}^{\pm \sigma }_{kl}}{\Lambda_k^0-\Lambda_l^0}  (\vec{\Psi}^\sigma_l)^0, 
   &
   \textbf{K}^{\sigma \nu }&=\frac{\Phi^\sigma \textbf{D} (\Phi^\nu)^T+\Psi^\sigma \textbf{D} (\Psi^\nu)^T}{2},
   & \sigma,\nu \in \{+,-\}.
 \end{align}
 The energy difference is smallest for $k,l$ close to $0,\frac{n}{2},n$. There we can expand the dispersion $\Lambda_k^0= \beta \phi_k^2 + \alpha \phi_k^4$ with $\beta=\frac{h-(1-\gamma ^2)}{2!|1-h|}$. At the nonequilibrium phase transition $h_c= 1-\gamma^2$ when $\beta = 0$, the next order has a prefactor $\alpha=\frac{3}{4!}(\frac{1}{\gamma^2}-1)$  .
 
 
 We can bound $| \textbf{K}^{\pm \alpha }_{kl} | <  \frac{\sum_j S_j}{n}$. This gives us the bound $n_b$ below which the perturbation theory is non-problematic:
 \begin{align}
 \left|  \frac{\textbf{K}^{\pm \alpha }_{kl}}{\Lambda_k^0-\Lambda_l^0} \right| & <\frac{\sum_j S_j}{n} \frac{1}{|\beta (\frac{2\pi}{n})^2 + \alpha  (\frac{2\pi}{n})^4|},
 & 
 n_b&= \frac{\beta (2\pi)^2}{\sum_j S_j} , & (n_b)_c&= \sqrt[3]{\frac{\alpha (2\pi)^4}{\sum_j S_j}} ,
 \end{align}
where the last bound $(n_b)_c$ is valid at the criticality when $\beta=0$. 
For larger $n$ we may need to use the degenerate perturbative theory at least for some energy levels and treat the non-exact degeneracy as a perturbation.

\subsection{Spectral Gap}
The spectral gap is an important feature of the system, since it sets the longest relaxation time. Looking at the smallest correction, we can calculate the spectral gap of the Liouvillian $\Delta= 2 \ \text{min}( \text{Re} \ \Lambda )=2 \ \text{min} \ \Lambda^{1}$.

The smallest gap appears close to $\phi_k=0,\pi$ ($k=1,n/2$), where for large $n$ holds that:
\begin{align}
    \lambda_1 &\approx  \frac{\gamma}{1-h} \frac{\pi}{n}, &&
    \lambda_{n/2} \approx -\frac{\gamma}{1+h} \frac{\pi}{n},
    \\
    \Delta_a&= \frac{S}{n^3} 2  \pi^2 \left(\frac{  \gamma}{1 \mp h} \right)^2, &&
    \Delta_b= \frac{S}{n^3} 4  \pi^2 \left(1+ \left(\frac{ \gamma}{1 \mp h} \right)^2\right),
\end{align}
where subscripts $a$ and $b$ refer to the cases a) and b) respectively. The case with $1+h$ results in lower energy. Notice the $n^{-3}$ dependence, which is the same as for open XY chain \cite{Prosen2008} (which has the same fermionic Hamiltonian up to periodic boundary term). 

This result is valid as long as we are away from the critical $h_c$ where the unperturbed dispersion is quartic, and holds for every large $n$, not just $n<n_b$, because: in the case a):
\begin{align}
  \textbf{K}^{- -}_{kl}&=\frac{S}{2n}\sin \lambda_k \sin \lambda_l,  & 
  \textbf{K}^{+ +}_{kl}&=\frac{S}{2n}\cos \lambda_k \cos \lambda_l, & 
  \textbf{K}^{\pm \mp}&=0.
\end{align}
The smallest eigenvalues's correction eigenvectors can couple only to $\textbf{K}^{- \alpha}$ and the correction to eigenvectors goes like $ \frac{\frac{S}{2n}\sin \lambda_k \sin \lambda_l}{\beta (\frac{2\pi}{n})^2} \approx \frac{S}{\beta n}$.

In the case b) the expressions become more complicated, so we state here only the dependencies on $n$ for $k,l$ close to 0
\begin{align}
  \textbf{K}^{- -}_{kl} &\approx \frac{S}{n^3}, & 
  \textbf{K}^{+ +}_{kl} &\approx \frac{S}{n}, & 
  \textbf{K}^{\pm \mp}_{kl} &\approx \frac{S}{n^2}.
\end{align}
Again the smallest eigenvalues's correction eigenvectors can couple only to $\textbf{K}^{- \alpha}$ and the largest correction 
is $ \frac{\textbf{K}^{-+}}{\beta (\frac{2\pi}{n})^2} \approx \frac{S}{\beta}$, so the expression for the NESS gap stays valid even for $n>n_b$, if we are not too close to the criticality where $\beta \to 0$.

\subsubsection{Spectral Gap at the Non-Equilibrium Phase Transition.}
At the non-equilibrium phase transition $\beta=0$ the energy levels are much closer together, therefore we can use the above arguments only up to the bound in the critical case $(n_b)_c$. 
After that we would need to treat more and more energy levels as degenerate and look at the almost degeneracy as a perturbation if we wished to continue with the perturbation theory approach.

\begin{figure}
    \centering
    \includegraphics[scale=.42]{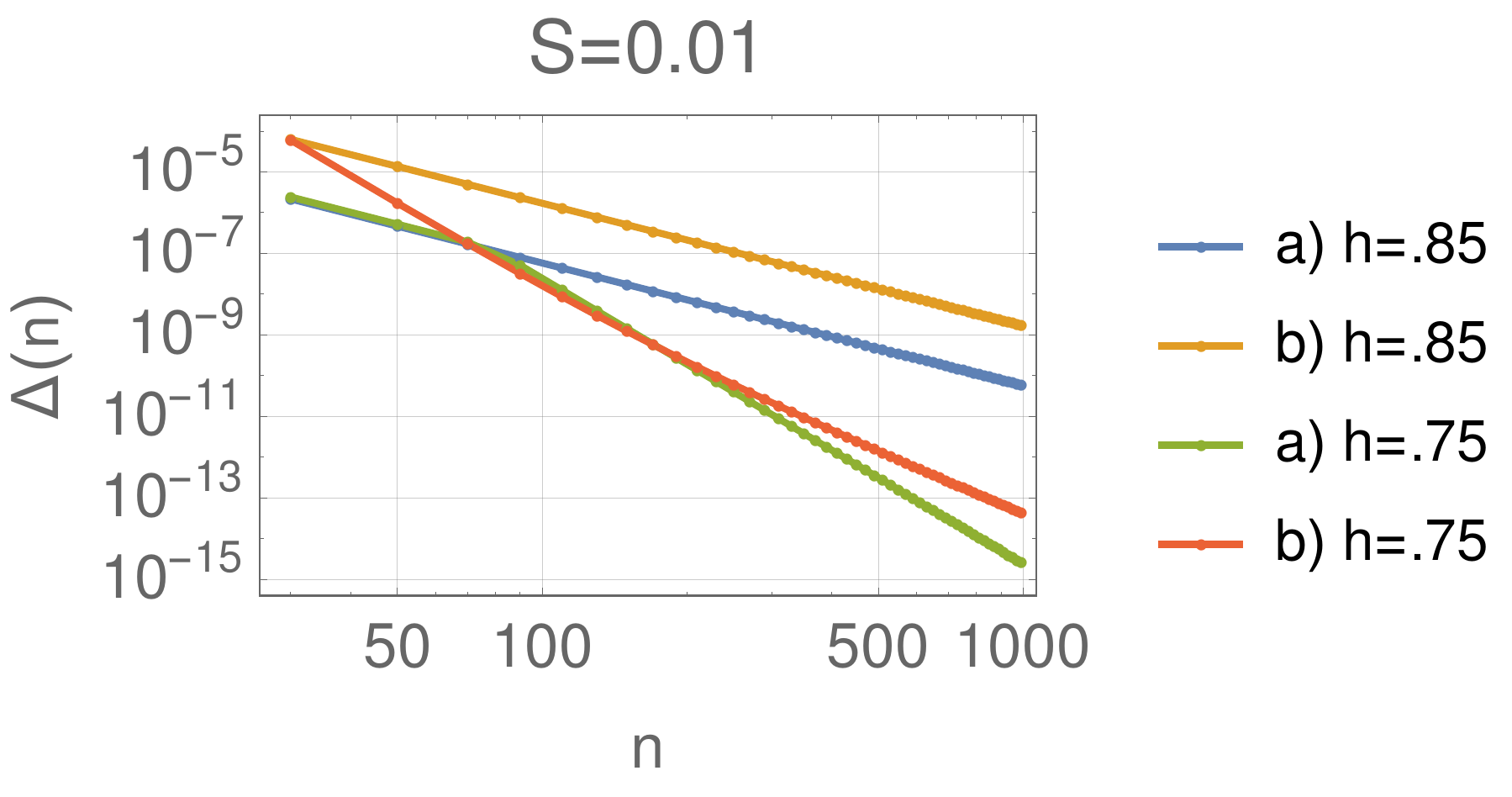}
    \includegraphics[scale=.42]{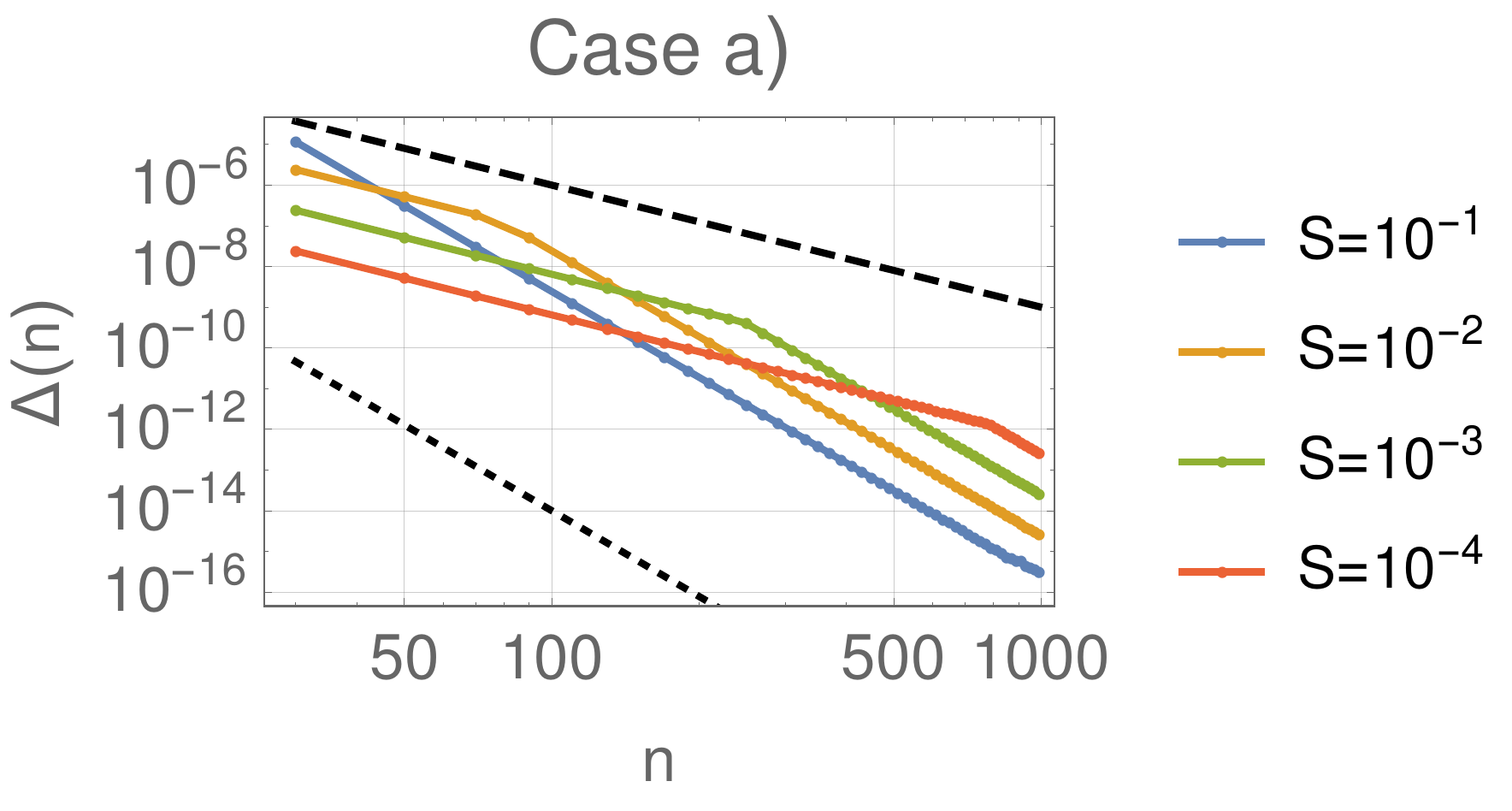}\\
    \includegraphics[scale=.47]{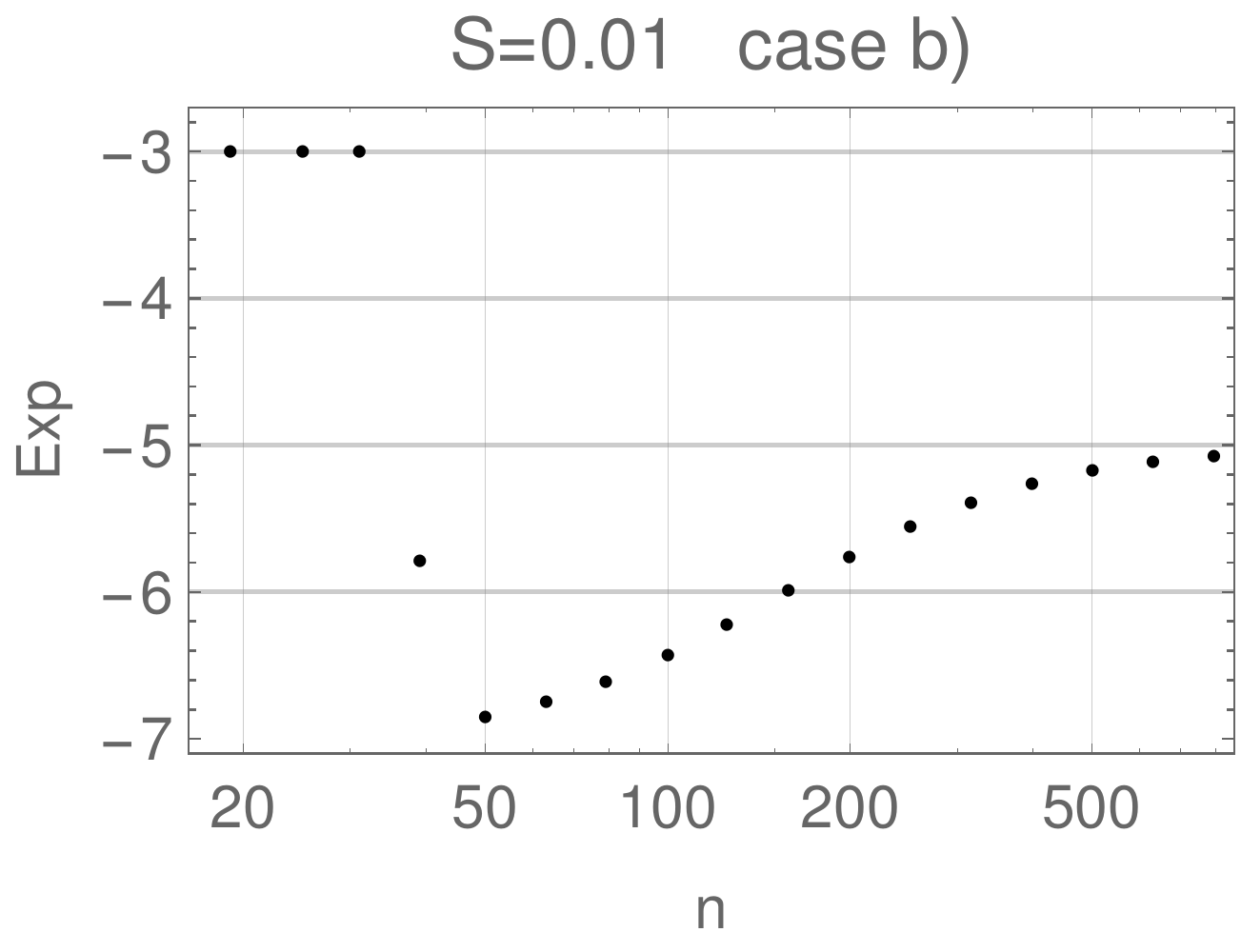}
\caption{
{\it Top left}: a comparison of the spectral gap between the critical and the non-critical regimes for cases a and b. In the critical regime (at the non-equilibrium phase transition $h_c=.75$), we see an interesting transition of the behaviour from $n^{-3}$ to $n^{-7}$ in the case a).
{\it Top right}: a lower coupling in case a) at the critical $h_c$ makes a transition from $n^{-3}$ to the $n^{-7}$ critical regime later, but in this regime a bigger counterintuitive prefactor $1/S$ emerges. The lines with slopes $-3$ and $-7$ are shown as dashed/dotted to guide the eye.
{\it Bottom}: the exponent in $n^x$, which we have calculated in the case b) numerically by interpolation at the system size $n$. It changes from $-3$ to $-7$ as in the case a), but then slowly changes to $-5$ as in the open XY spin chain. In all figures we set $\gamma=0.5$.
}
\label{fig:ring_gap}
\end{figure}

The numerical results show an interesting behaviour that is summarized in figure \ref{fig:ring_gap}. The gap size scaling in $n$ changes from $S n^{-3}$ to $n^{-7}/S$ in the case a). It is interesting that the gap closes faster for larger couplings $S$, which is counterintuitive.

In the case b) we make a transition from $n^{-3}$ to $n^{-7}$ and for very large $n$ to $n^{-5}$, the latter is the same as in the open XY spin chain \cite{Prosen2008}.
%

\section{Discussion and Conclusions}

We have discussed open fermionic systems with a quadratic Hamiltonian and a linear dissipation governed by the Lindblad equation and written a simple equation \eqref{eq:dynamics1} governing the dynamics of the system after a local quench. The driving 
only influences the correlations in the NESS, whereas the dissipation 
governs the decay to the NESS and the Hamiltonian gives the oscillatory response.

Looking at the response to local quench in the open XY spin chain driven at the edges, which can be mapped to our problem, reveals a light cone behaviour, which is asymmetric for the driving at different effective temperatures. The dynamic response of the two dimensional problem behaves similarly. 
Another interesting example is the XY spin chain with Dzyaloshinskii-Moriya interactions, where we have noticed a new type of non-equilibrium phase transition with the long-range correlated NESS which is due to the appearance of a pair of Dirac points in the dispersion relation.

In the second part of the paper, we have presented an analytic solution for a weakly coupled fermionic ring with nearest-neighbour hopping, superconductive and chemical potential terms in the Hamiltonian. The explicit real corrections to the imaginary eigenvalues and the correlations in the NESS 
are calculated based on the solution of the closed XY spin ring, which is perturbed by a weak coupling.

The special case of dissipative driving at just a single site gives a very simple and highly non-local NESS, where each pair of opposite sites is strongly correlated (entangled).  We mention the non-equilibrium phase transition which gives long-range correlations in the NESS (that was reported for open spin XY model \cite{Prosen2008a}) by looking at the spectral gap and the two point correlation functions in the NESS. Interestingly, to notice the effect with the long-range correlations, we need driving at more than one site.

There remain many interesting open questions.
One can easily numerically treat general quadratic driven systems, so one could look at the far from equilibrium dynamics in many other systems, for example in the disordered systems, the three dimensional systems or the long-range coupled systems. This framework can also be a starting point for the perturbation theory of the non-quadratic (weakly interacting) systems.

\section*{Acknowledgments}
P.K. thanks Enej Ilievski, Berislav Bu\v{c}a and Spyros Sotiriadis for helpful discussions and  Bojan \v{Z}unkovi\v{c} for helpful discussions and comments on the manuscript. The work has been supported with the Grant N1-0025 of Slovenian Research Agency (ARRS) and Advanced grant OMNES of European Research Council (ERC).

\section*{References}
\nocite{*}
\bibliographystyle{unsrt}
\bibliography{references}

\end{document}